\documentclass[oneside,reqno]{amsart}
\usepackage{amsmath, amsthm, mathrsfs, amssymb, amsfonts, mathtools}
\usepackage{graphicx}
\usepackage[dvipsnames]{xcolor}
\usepackage{hyperref}
\usepackage[backend=biber, style=phys, doi=false, isbn=false, hyperref=true, backref=false, firstinits=true, abbreviate=true, sorting=nyt]{biblatex}
\bibliography{lib}
\def\diff{\mathrm{d}}

\DeclareMathOperator{\Ci}{Ci}
\DeclareMathOperator{\Si}{Si}
\newcommand{\inner}[2]{\left(\left.{#1}\right|{#2}\right)}
\DeclareMathOperator{\re}{Re}

\usepackage[foot]{amsaddr}
\begin{document}

\title{Resonant Dynamics and the Instability of the Box Minkowski model}

\author{Jo\"el Kurzweil$^1$}
\email[]{a1308608@unet.univie.ac.at}
\author{Maciej Maliborski$^{1,2,\dagger}$}
\address{$^1$University of Vienna, Gravitational Physics, Boltzmanngasse 5, 1090 Vienna, Austria}
\address{$^2$University of Vienna, Faculty of Mathematics, Oskar-Morgenstern-Platz 1, 1090 Vienna, Austria}
\email[]{maciej.maliborski@univie.ac.at}
\address{$^{\dagger}$Corresponding author}

\begin{abstract}

We revisit the \textit{box Minkowski model} \cite{Maliborski.2012} and provide a strong argument that, subject to the Dirichlet boundary condition, it is unstable toward black hole formation for arbitrarily small generic perturbations. Using weakly nonlinear perturbation theory, we derive the resonant system, which compared to systems with the anti-de Sitter asymptotics, has extra resonant terms, and study its properties, including conserved quantities. We find that the generic solution of the resonant system becomes singular in finite time. Surprisingly, the additional resonant interactions do not significantly affect the singular evolution. Furthermore, we find that the interaction coefficients take a relatively simple form, making this a particularly attractive toy model of turbulent gravitational instability.

\end{abstract}

\date{\today}

\maketitle

\tableofcontents

\section{Introduction}

Over the past several years, research on asymptotically anti-de Sitter (AdS) spacetimes, in particular regarding their stability, has greatly intensified. This surge was triggered by \cite{Bizoń.2011} which demonstrated, using a direct numerical solution of the spherically symmetric Einstein-Scalar field system with negative cosmological constant, together with a scaling argument, that the AdS solution is unstable toward black hole formation for generic arbitrarily small initial perturbations.

Subsequent works, see \cite{Evnin.2021v} for a recent review, identified the key components for this instability. Namely, the confinement realized by a suitable choice of boundary condition at the conformal boundary, and related to that, the resonant spectrum of linear perturbations (see below for a definition).

Of the many works which appeared over the years on that topic, some of the most influential were \cite{Balasubramanian.2014,Craps.2014,Craps.2015}. Those studies extended the initial perturbative calculations of \cite{Bizoń.2011} and analyzed the resonant interactions between the linear modes to a greater extent. Incidentally, this perturbative technique not only improved the perturbative expansion of \cite{Bizoń.2011}, which was prone to the secular terms identified as the progenitors for the instability, but it also provided yet another piece of evidence for the instability of AdS \cite{Bizoń.2015, Deppe.20191vp} and thus strengthened the scaling argument presented in \cite{Bizoń.2011}.

In the quest to further understand spatially confined systems and, in particular, to cast new light on the stability of gravitational dynamics, we revise the \textit{box Minkowski model} initially studied in \cite{Maliborski.2012}. 
Using the resonant approach \cite{Bizoń.2015}, we investigate the future evolution of small spherically symmetric scalar perturbations of the flat space subject to Dirichlet boundary conditions imposed at the perfectly reflecting cavity located at a finite radius. Our findings strengthen the previous results \cite{Maliborski.2012} and provide evidence that the model is unstable toward black hole formation for generic arbitrarily small initial data. This study is yet another demonstration that the confinement and the resonant spectrum are indeed the key components for turbulent instability, though not sufficient.

Although the model's nongeometric origin might not appeal to some, we argue that this might be one of the simplest models that exhibit the turbulent dynamics observed in gravitating systems \cite{Bizoń.2011, Bizoń.2017}. This is the case despite some of the most striking differences compared to AdS, namely the existence of extra resonant interaction channels as well as the restricted symmetry of the coefficients of the resonant system. 

We view the \textit{box Minkowski model} as an attractive toy model for the corresponding problem with a negative cosmological constant. Our results indicate that the perturbative approximation correctly captures the energy transfer between the eigenmodes of the linearized problem and suggest that the perturbative solution is a good approximation of the nonlinear solution up to apparent horizon formation. Additionally, this study may shed new light on the elusive nature of the singular solution of the resonant system in AdS$_{4}$ \cite{Deppe.20191vp}. 

The manuscript is structured as follows. In Sec.~\ref{sec:TheModel}, we introduce the model and review its most important features relevant for this work. In Sec.~\ref{sec:ResonantApproximation}, we present the derivation of the resonant approximation and discuss its properties (interaction coefficients and conserved quantities). The numerical and asymptotic study of the solutions of the resonant system and comparison with the Einstein-Scalar field system are presented Sec.~\ref{sec:Results}. We conclude in Sec.~\ref{sec:Conclusions}. In the Appendix~\ref{sec:BoundaryTimeGauge}, we discuss the effect the residual time gauge has on the singular solution.

\section{The model}
\label{sec:TheModel}

\subsection{Equations}

We study the \textit{box Minkowski model} of \cite{Maliborski.2012}, which is a minimally coupled self-gravitating massless scalar field with zero cosmological constant in four spacetime dimensions,
\begin{equation}
	\label{eq:22.04.21_01}
	G_{\mu\nu} = 8\pi G \left(\nabla_{\mu}\phi\nabla_{\nu}\phi - \frac{1}{2}g_{\mu\nu}g^{\alpha\beta}\nabla_{\alpha}\phi\nabla_{\beta}\phi\right)
	\,,
	\quad
	g^{\mu\nu}\nabla_{\mu}\nabla_{\nu}\phi = 0
	\,.
\end{equation}
To mimic the reflecting boundary condition of the asymptotically AdS case \cite{Bizoń.2011}, we require the evolution to be confined in a perfectly reflecting spherical cavity of fixed radius $r=R>0$. Furthermore, we assume spherical symmetry, so the solution outside the cavity is that of a Schwarzschild spacetime, with a mass parameter equal to the total mass of the interior configuration.
We do not consider the issue of smoothness of the solution across the boundary,\footnote{The global solution, i.e. solution which extends to all radii, is continuous but not differentiable at $r=R$.} neither do we discuss the issue of a potential physical realization of such a model. 

For clarity of presentation, we state the field equations as they appear in \cite{Maliborski.2012}. We use the following ansatz of a spherically symmetric asymptotically flat metric written in spherical polar coordinates $(r,\theta,\varphi)$:
\begin{equation}
	\label{eq:20.12.18_01}
	\diff{s}^{2} = -Ae^{-2\delta}\diff{t}^{2} + A^{-1}\diff{r}^{2} 
	+ r^{2}\left(\diff{\theta}^{2} + \sin^{2}{\theta}\diff{\varphi}^{2}\right)\,,
\end{equation}
where the metric functions $A$ and $\delta$ depend on $t$ and $r$ only. The equations of motion \eqref{eq:22.04.21_01} are
\begin{align}
	\label{eq:20.12.18_02}
	\delta' &= -r\left(\Phi^{2} + \Pi^{2}\right)\,,
	\\
	\label{eq:20.12.27_03}
	A' &= \frac{1-A}{r} -rA\left(\Phi^{2} + \Pi^{2}\right)\,,
	\\
	\label{eq:20.12.18_03}
	\dot{\phi} &= \Pi Ae^{-\delta}\,,
	\\
	\label{eq:20.12.18_04}
	\dot{\Pi} &= \frac{1}{r^{2}}\left(r^{2}Ae^{-\delta}\phi'\right)'\,,
\end{align}
where we introduced auxiliary variables $\Phi\equiv\phi'$ and $\Pi\equiv e^{\delta}\dot{\phi}/A$. We adopt the notation where $\dot{}$ and $'$ stand for time and radial derivatives, respectively. We use the units where $c=1$ and $4\pi G=1$. Without loss of generality, we set $R=1$ (which can always be obtained by a suitable rescaling of time and radial coordinates).

We want to study the future evolution of small initial data subject to the reflective boundary condition at the cavity. These follow from the consideration of the mass of the system. The total mass of the system
\begin{equation}
	\label{eq:22.06.21_04}
	M = \frac{1}{2}\int_{0}^{1}A\left(\Phi^{2} + \Pi^{2}\right)r^{2}\diff{r}
	\,,
\end{equation}
is constant whenever $\Phi\Pi|_{r=1}=0$ is satisfied, c.f., \cite{Maliborski.2015}. Then there is no energy flux through the boundary. 
Below we motivate our choice of the Dirichlet condition $\phi|_{r=1}=0$.

\subsection{Linear problem}
\label{sec:LinearProblem}
First, we look at linear perturbations of the vacuum solution $\phi\equiv 0$, $A=1$, $\delta=0$. In this case, \eqref{eq:20.12.18_02}-\eqref{eq:20.12.18_04} reduce to a free wave equation in spherical symmetry
\begin{equation}
	\label{eq:20.12.18_05}
	\ddot{\phi} + L\phi = 0\,,
	\quad
	L = -\frac{1}{r^{2}}\partial_{r}\left(r^{2}\partial_{r}\right)\,,
\end{equation}
with Dirichlet condition at the cavity $\phi|_{r=1}=0$. The eigenvalues and eigenfrequencies of the operator $L$ are
\begin{equation}
	\label{eq:20.12.18_06}
	e_{j}(r) = \sqrt{2}\,\frac{\sin{\omega_{j}r}}{r}\,,
	\quad
	\omega_{j} = (j+1)\pi\,, \quad j=0,1,\ldots
	\,.
\end{equation}
The functions $e_{j}(r)$ form an orthonormal basis on the Hilbert space $L^{2}\left([0,1],r^{2}\diff{r}\right)$ with respect to the inner product
\begin{equation}
	\label{eq:22.06.21_02}
	\left(\xi\,|\chi\right) =
	\int_{0}^{1}\xi(r)\chi(r)r^{2}\diff{r}
	\,.
\end{equation}

As for the linear perturbations of AdS space \cite{Bizoń.2011}, the spectrum \eqref{eq:20.12.18_06} is completely resonant. This means that the eigenfrequencies are rational multiples of one another, which then implies that for any $i,j,k=0,1,\ldots$, a combination $\omega_{i}\pm\omega_{j}\pm\omega_{k}$ is (modulo sign) also an eigenfrequency. This fact and the nonlinearities of the governing equations lead to resonant interactions between modes, which result in complex, turbulent dynamics \cite{Bizoń.2011}. To capture these interactions, we use a weakly nonlinear expansion, the main topic of the following sections.

We note that for the boundary condition $\partial_{r}\phi|_{r=1}=0$, for which the total mass \eqref{eq:22.06.21_04} is also conserved, the eigenfrequencies are only asymptotically resonant, i.e. $\omega_{j}\sim\pi(j+1/2)$ for $j\rightarrow\infty$. Numerical data suggests that the dispersion introduced by the non-resonant spectrum obstructs the collapse of very small initial data \cite{Maliborski.2014, Maliborski.2015}.
On the perturbative level, there are no couplings between modes for non-resonant eigenfrequencies, as the resonant system is trivial. Of course, there still could be self-interactions. However, they merely affect mode phases, but not their amplitudes. This supports the observation that no black hole forms, at least not at the time scale $\varepsilon^{-2}$, where $\varepsilon\rightarrow 0$ measures the size of the initial perturbation.
The model can be extended to higher dimensions, but the linear spectrum is resonant in four spacetime dimensions and for the Dirichlet boundary condition only. The other choices lead to the asymptotically resonant eigenfrequencies. This motivates the study of the particular case we consider here, as it corresponds to the completely resonant spectrum of scalar perturbations of AdS \cite{Bizoń.2011}.

Note that a solution to the initial-boundary value problem introduces corner conditions at the cavity. In order to guarantee a smooth evolution, certain relations on the coefficients of the Taylor expansion at the time-like boundary $r=1$ need to be satisfied, see \cite{Maliborski.2015}. This, in turn, restricts permissible initial data. In particular, any finite combination of eigenfunctions \eqref{eq:20.12.18_06} violates these conditions, as verified in \cite{Maliborski.2015}.
Therefore, in this work, we consider initial data with sufficient decay as $r\rightarrow 1$ so that the corner conditions are automatically satisfied.

\subsection{Nonlinear evolution}

The solution to the initial-boundary value problem was already presented in \cite{Maliborski.2012, Maliborski.2014}, see also \cite{Maliborski.2015}. Here we briefly review those results which are essential for the subsequent analysis.

We solved the Einstein-Scalar field equations and the resonant system for various \linebreak ``Gaussian-like'' initial conditions and observed a similar behavior within this class of data, i.e. turbulent evolution leading to the growth of the Ricci scalar and the development of polynomial spectra of mode energies with a universal exponent.\footnote{We expect that there exist initial conditions which belong to islands of stability spanned by the time-periodic solutions of the model, constructed in \cite{Maliborski.2015}. Such solutions avoid the collapse at least at the timescale $\varepsilon^{-2}$ \cite{Maliborski.2013rhg, Maliborski.2014}.} For clarity of presentation, we focus on the particular choice
\begin{equation}
	\label{eq:20.12_01}
	\Phi(0,r) = 0\,, \quad \Pi(0,r) = \varepsilon\exp\left(-64\tan^{2}\left(\frac{\pi}{2}r\right)\right)\,,
\end{equation}
and we only present the evidence that the instability is generic by considering also
\begin{equation}
	\label{eq:20.12_01b}
	\phi(0,r) = \varepsilon \exp\left(-\left((r-1)^{-2}r^{-2}-16\right)\right)\,, \quad \Phi(0,r) = \partial_{r}\phi(0,r)\,, \quad \Pi(0,r) = \partial_{r}\phi(0,r)
	\,,
\end{equation}
see Fig.~\ref{fig:RicciOriginEinsteinVSResonant} below. In both cases, $\varepsilon$ is a small parameter.

\begin{figure}
	\includegraphics[width=0.45\textwidth]{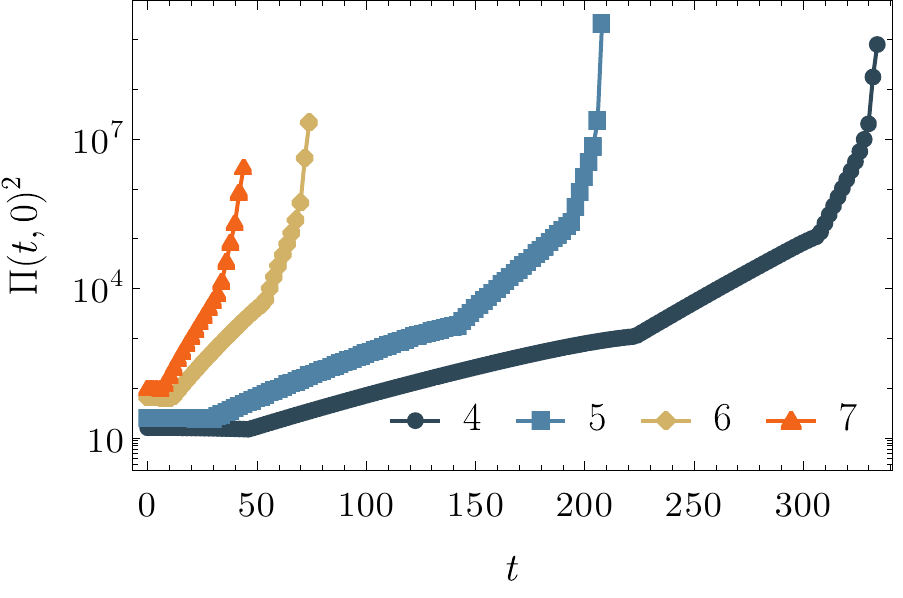}
	\hspace{1ex}
	\includegraphics[width=0.45\textwidth]{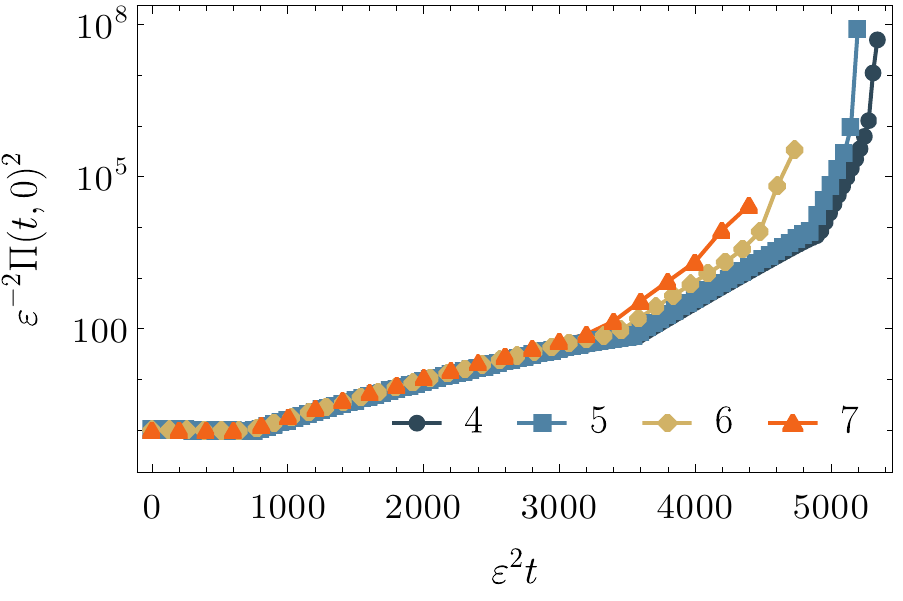}
	\caption{The upper envelope of the Ricci scalar evaluated at the origin for solutions with different sizes of initial perturbation in \eqref{eq:20.12_01}. The right plot shows the rescaled $\varepsilon^{-2}\Pi(\varepsilon^{2}t,0)^{2}$ function.}
	\label{fig:RicciScalar}
\end{figure}

The solution of the initial-boundary value problem \eqref{eq:20.12.18_02}-\eqref{eq:20.12.18_04} uses techniques described in detail in \cite{Maliborski.2012, Maliborski.2014, Maliborski.2015}, see also \cite{Maliborski.2013}. The code is an improved, and parallelized version used in \cite{Maliborski.2012, Maliborski.2015}. It is based on the method of lines with a fourth-order spatial finite-difference discretization scheme. For time integration, we take the fourth-order classical Runge-Kutta method. The time step $\Delta t$ is adjusted so that $1/6\leq\Delta t/\Delta r e^{-\delta_{\text{max}}}\leq 1/3$, $\delta_{\text{max}}\equiv\max_{r}\delta(t,r)$, for fixed spatial resolution $\Delta r$. Typical spatial resolutions vary from $2^{16}$ to $2^{18}$ uniform grid points depending on the amplitude of the initial data (lower amplitudes require finer grid spacing to resolve steep gradients and for the data to collapse; the apparent horizon is indicated by the minimum of the metric function $A$ dropping below $2^{3-k/2}$ on grids with $2^{k}$ points). We add the Kreiss-Oliger type artificial dissipation to filter out high frequencies.  The code was demonstrated to be fourth-order convergent and highly accurate in following the solution up to a black hole formation \cite{Maliborski.2015}.

\begin{figure}
	\includegraphics[width=0.5\textwidth]{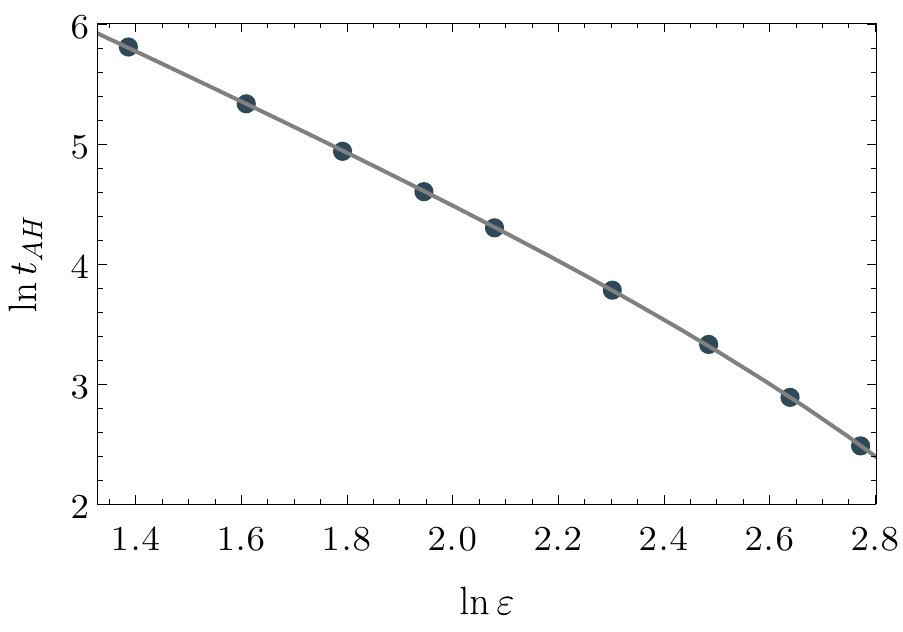}
	\caption{The apparent horizon formation time $t_{AH}$ as a function of the amplitude $\varepsilon$ in the initial data \eqref{eq:20.12_01}. The solid line shows the fit $\ln t_{AH}(\varepsilon) = -2 \ln \varepsilon + a + b\varepsilon^{2}$. From this we find $\tau_{AH} = \lim_{\varepsilon\rightarrow 0} \varepsilon^{2} t_{AH}(\varepsilon)=\exp(a)\approx 5499.02$.}
	\label{fig:TimeAHvsAmplitude}
\end{figure}

By decreasing the magnitude of the initial data, we observe the onset of instability and eventual formation of an apparent horizon (AH) at later times.\footnote{The AH formation is indicated by the metric function $A$ dropping to zero.} The scaling of the Ricci scalar evaluated at the coordinate origin suggests that $\varepsilon^{-2}$ is the instability timescale, see Fig.~\ref{fig:RicciScalar}. Indeed, the fit to the AH formation time $t_{AH}(\varepsilon)$ presented in Fig.~\ref{fig:TimeAHvsAmplitude} strongly suggests that the limit
\begin{equation}
	\label{eq:22.07.12_01}
	\tau_{AH} := \lim_{\varepsilon\rightarrow 0} \varepsilon^{2} t_{AH}(\varepsilon)
\end{equation}
exists and is finite. For the considered initial data \eqref{eq:20.12_01}, we find
\begin{equation}
	\label{eq:22.07.12_02}
	\tau_{AH} \approx 5499.02
	\,.
\end{equation}
Later, we will see that this limit agrees well with the prediction obtained from the resonant approximation.

Moreover, to quantify the energy transfer between the modes and to investigate the solution close to the AH formation, we rewrite the total mass \eqref{eq:22.06.21_04} as the Parseval sum
\begin{equation}
	\label{eq:22.07.12_03}
	M = \sum_{j\geq 0}E_{j}(t)
	\,,
\end{equation}
where
\begin{equation}
	\label{eq:22.06.21_01}
	E_{j} = \Pi_{j}^{2} + \omega_{j}^{-2}\Phi_{j}^{2}\,, \quad \Phi_{j}=\left(\left.A^{1/2}\Phi\,\right|e_{j}'\right), \quad \Pi_{j}=\left(\left.A^{1/2}\Pi\,\right|e_{j}\right)\,,
\end{equation}
and $E_{j}$ can be interpreted as the energy contained in the linear mode $e_{j}$. 
\begin{figure}
	\includegraphics[width=0.45\textwidth]{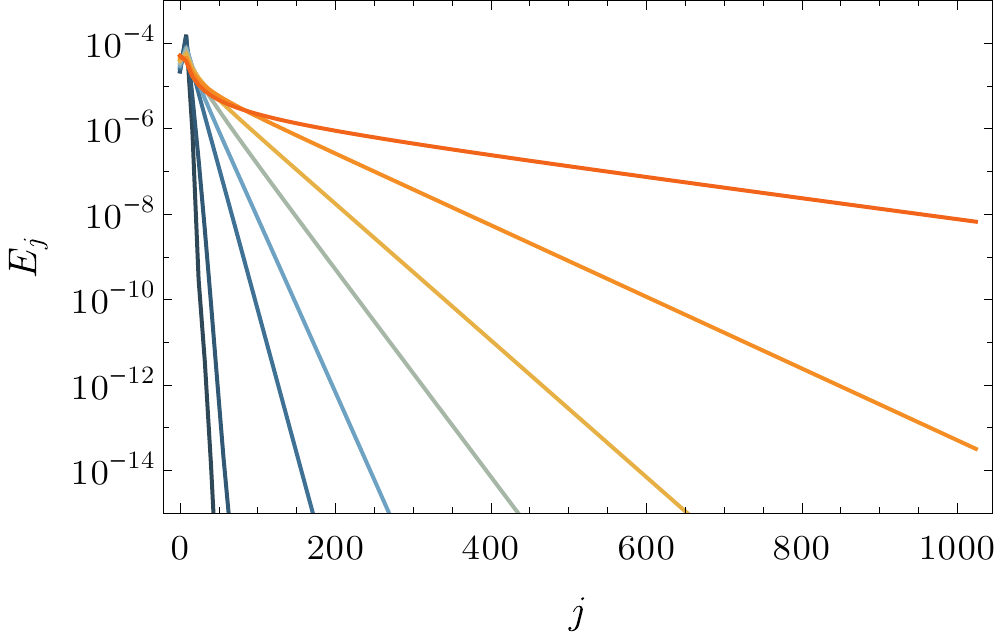}
	\hspace{1ex}
	\includegraphics[width=0.45\textwidth]{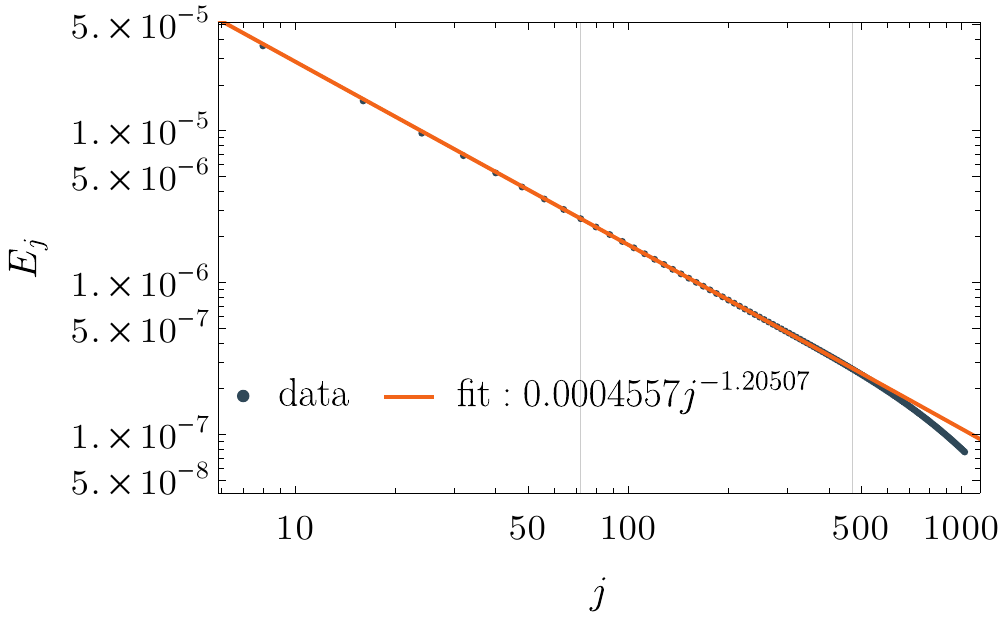}
	\caption{Left: the time development of the energy spectra during the collapse. As time progresses, the energy \eqref{eq:22.06.21_01} is shifted to higher modes. Right: in the late phases of the evolution, before the AH is detected, the energy exhibits the power law $E_{j}\sim j^{-\mu}$. The fit gives $\mu\approx 1.2$, which is close to the exponent found in the study of AdS$_{4}$. We present data with the smallest amplitude considered $\varepsilon=4$.}
	\label{fig:EnergySpectra}
\end{figure}
Close to the time of AH formation, we see the development of a power law spectrum
\begin{equation}
	\label{eq:22.07.12_04}
	E_{j}(t\approx t_{AH}) \sim j^{-\mu}, \quad \mu \approx 6/5
	\,,
\end{equation}
see Fig.~\ref{fig:EnergySpectra}. This indicates that the solution loses smoothness during the collapse. The value of the exponent in \eqref{eq:22.07.12_04} appears universal, independent of an initial perturbation, for a class of data exhibiting scaling as in \eqref{eq:22.07.12_01}. We note that a similar exponent was observed in the AdS$_{4}$ case \cite{Bizoń.2011}. However, the exponent depends both on the spacetime dimension and on the particular `matter' model \cite{Maliborski.2013, Bizoń.2017}. Up to now, there is no satisfactory explanation or derivation of its value.\footnote{For the Einstein-Scalar field-AdS model in $(d+1)$ spacetime dimensions the dimensional argument of Bizo\'n and Rostworowski \cite{Bizoń.2022} predicts the energy spectra $E_{j}\sim j^{-(d-2)}$, but cf. \cite{Maliborski.2013} for another suggestion which is based on numerical results performed for AdS in different dimensions.}

We remark that these results were obtained for the origin time gauge, where $t$ is the proper time of the centre observer. We repeated the analysis using the data obtained for the alternative boundary time gauge in which the time coordinate $t$ corresponds to the proper time of the observer located at $r=1$. Although defining the AH formation time is more difficult in that case, due to the redshift effect as the AH is about to form, we obtained very similar results by tweaking the threshold for black hole formation. In particular, the extrapolation as in \eqref{eq:22.07.12_01} gives $\tau_{AH}\approx 5484.59$.

\section{Resonant approximation}
\label{sec:ResonantApproximation}

\subsection{Resonant system}
The derivation of the resonant approximation closely follows \cite{Craps.2014,Craps.2015}. We start with the standard perturbative expansion, as in \cite{Craps.2014,Craps.2015}, but contrary to these works, we arrive at the resonant system by using multi-scale analysis \cite{Murdock.1999book} as it was done in \cite{Balasubramanian.2014}.

We start with the perturbative ansatz
\begin{align}
	\label{eq:20.12.18_08}
	\phi(t,r) &= \varepsilon \phi_{1}(t,r) + \varepsilon^{3}\phi_{3}(t,r) + \cdots\,,
	\\
	\label{eq:20.12.18_09}
	A(t,r) &= 1- \varepsilon^{2} A_{2}(t,r) + \cdots\,,
	\\
	\label{eq:20.12.18_10}
	\delta(t,r) &= \varepsilon^{2}\delta_{2}(t,r) + \cdots\,,
\end{align}
where $\varepsilon$ is a small parameter controlling the size of a perturbation of flat space $\phi\equiv 0$, $A\equiv 1$ and $\delta\equiv 0$. We plug in the series \eqref{eq:20.12.18_08}-\eqref{eq:20.12.18_10} into \eqref{eq:20.12.18_02}-\eqref{eq:20.12.18_04} expand around $\varepsilon=0$ and set to zero terms with equal powers of $\varepsilon$. As a result, we get a sequence of differential equations that we solve order by order. At the leading order, we find
\begin{equation}
	\label{eq:20.12.18_11}
	\ddot{\phi}_{1} + L\phi_{1} = 0\,,
\end{equation}
cf. \eqref{eq:20.12.18_05}, thus a generic solution satisfying $\phi_{1}(t,r=1)=0$ can be written as
\begin{equation}
	\label{eq:20.12.18_12}
	\phi_{1}(t,r) = \sum_{j\geq 0}c_{j}(t)e_{j}(r)\,,
\end{equation}
where
\begin{equation}
	\label{eq:21.08.02_01}
	c_{j}(t) = \alpha_j e^{i\omega_{j}t} + \bar{\alpha}_{j} e^{-i\omega_{j}t}\,,
\end{equation}
and $\alpha_{j}$ are constant parameters uniquely determined by the initial data $\phi|_{t=0}=\varepsilon f(r)$ and $\partial_{t}\phi|_{t=0}=\varepsilon p(r)$.

At the second order, we get the equations
\begin{align}
	\label{eq:20.12.27_02}
	A_{2}' &= -\frac{A_{2}}{r} + r \left( \left(\phi_{1}' \right)^2 + \left(\dot{\phi}_{1} \right)^2 \right)
	\,,
	\\
	\label{eq:20.12.27_01}
	\delta_{2}' &= -r \left( \left(\phi_{1}' \right)^2 + \left(\dot{\phi}_{1} \right)^2 \right)
	\,, 
\end{align}
with respective solutions
\begin{align}
	\label{eq:20.12.18_29}
	A_{2}(t,r) &= \frac{1}{r}\int_{0}^{r}\diff{s}s^{2}\left(\phi_{1}'(t,s)^{2} + \dot{\phi}_{1}(t,s)^{2}\right)\,,
	\\
	\label{eq:20.12.18_30}
	\delta_{2}(t,r) &= - \int_{0}^{r}\diff{s}s\left(\phi_{1}'(t,s)^{2} + \dot{\phi}_{1}(t,s)^{2}\right)
	\,.
\end{align}
Combining \eqref{eq:20.12.27_02} and \eqref{eq:20.12.27_01}, we get the identity
\begin{equation}
	\label{eq:20.12.18_31}
	A_{2}' + \delta_{2}' = -\frac{1}{r^{2}}\int_{0}^{r}\diff{s}s^{2}\left(\phi_{1}'(t,s)^{2} + \dot{\phi}_{1}(t,s)^{2}\right)
	\,,
\end{equation}
which we will use at the later stages of the derivation.

Note that here we use the residual gauge freedom and set $\delta(t,0)=0$ so that the coordinate time $t$ is the proper time of the central observer. The gauge $\delta(t,1)=0$ is discussed in Appendix~\ref{sec:BoundaryTimeGauge}.

At third order, we find
\begin{equation}
	\label{eq:20.12.18_15}
	\ddot{\phi}_{3} + L\phi_{3} = S_{3} \equiv -(A_{2}'+\delta_{2}')\phi_{1}' 
	- (\dot{A}_{2}+\dot{\delta}_{2})\dot{\phi}_{1} 
	- 2(A_{2}+\delta_{2})\ddot{\phi}_{1}
	\,.
\end{equation}
We solve \eqref{eq:20.12.18_15} by decomposing $\phi_{3}$ in terms of eigenbasis \eqref{eq:20.12.18_06}:
\begin{equation}
	\label{eq:20.12.18_19}
	\phi_{3}(t,r) = \sum_{j\geq 0} c^{(3)}_{j}(t)e_{j}(r)
	\,,
\end{equation}
cf. \eqref{eq:20.12.18_12}.

Using \eqref{eq:20.12.18_19} and projecting \eqref{eq:20.12.18_15} onto the mode $e_{l}$ we get a system of coupled differential equations for the mode coefficients $c^{(3)}_{l}$
\begin{equation}
	\label{eq:20.12.18_28}
	\ddot{c}^{(3)}_{l} + \omega_{l}^{2}c^{(3)}_{l} = \left(\left.S_{3}\right|e_{l}\right)
	\,.
\end{equation}
In order to compute the projection in \eqref{eq:20.12.18_28}, we work out each term on the right-hand side (rhs) of \eqref{eq:20.12.18_28} separately and combine the results later. We find
\begin{align}
	\label{eq:20.12.18_32}
	\inner{\left(A_{2}' + \delta_{2}'\right)\phi_{1}'}{e_{l}} &= -\int_{0}^{1} \diff{r} r^{2} e_{l}(r) \phi_{1}'(t,r)\frac{1}{r^{2}} \int_{0}^{r} \diff{s} s^{2} \left(\phi_{1}'(t,s)^{2} + \dot{\phi}_{1}(t,s)^{2}\right) \notag \\
	&= -\sum_{i,j,k=0}^{\infty} c_{k}(t) \int_{0}^{1} \diff{r} e_{l}(r) e_{k}'(r) \notag \\
	&\quad \int_{0}^{r} \diff{s} s^{2} \left[ c_{i}(t) c_{j}(t) e_{i}'(s) e_{j}'(s)  + \dot{c}_{i}(t) \dot{c}_{j}(t) e_{i}(s) e_{j}(s)\right]\,,
\end{align}

\begin{align}
	\label{eq:20.12.18_33}
	\inner{\dot{A}_{2}\dot{\phi}_{1}}{e_{l}} &= \int_{0}^{1} \diff{r} r^{2} e_{l}(r) \dot{\phi}_{1}(t,r) \frac{1}{r} \int_{0}^{r} \diff{s} s^{2} \left(\frac{\partial}{\partial t} \phi_{1}'(t,s)^{2} + \frac{\partial}{\partial t} \dot{\phi}_{1}(t,s)^{2}\right) \notag \\
	&= \sum_{i,j,k=0}^{\infty} \dot{c}_{k}(t) \int_{0}^{1} \diff{r} r e_{l}(r) e_{k}(r) \notag \\
	&\quad \int_{0}^{r} \diff{s} s^{2} \left[\frac{\partial}{\partial t} \left(c_{i}(t) c_{j}(t) \right)e_{i}'(s) e_{j}'(s) +  \frac{\partial}{\partial t} \left(\dot{c}_{i}(t) \dot{c}_{j}(t) \right)e_{i}(s) e_{j}(s)\right]\,,
\end{align}

\begin{align}
	\label{eq:20.12.18_34}
	\inner{\dot{\delta}_{2}\dot{\phi}_{1}}{e_{l}} &= -\int_{0}^{1} \diff{r} r^{2} e_{l}(r) \dot{\phi}_{1}(t,r) \int_{0}^{r} \diff{s} s \left(\frac{\partial}{\partial t} \phi_{1}'(t,s)^{2} + \frac{\partial}{\partial t} \dot{\phi}_{1}(t,s)^{2}\right) \notag \\
	&= -\sum_{i,j,k=0}^{\infty} \dot{c}_{k}(t) \int_{0}^{1} \diff{r} r^{2} e_{l}(r) e_{k}(r) \notag \\
	&\quad \int_{0}^{r} \diff{s} s \left[\frac{\partial}{\partial t} \left(c_{i}(t) c_{j}(t) \right)e_{i}'(s) e_{j}'(s) +  \frac{\partial}{\partial t} \left(\dot{c}_{i}(t) \dot{c}_{j}(t) \right)e_{i}(s) e_{j}(s)\right]\,,
\end{align}

\begin{align}
	\label{eq:20.12.18_35}
	\inner{A_{2}\ddot{\phi}_{1}}{e_{l}} &= \int_{0}^{1} \diff{r} r^{2} e_{l}(r) \ddot{\phi}_{1}(t,r) \frac{1}{r} \int_{0}^{r} \diff{s} s^{2} \left(\phi_{1}'(t,s)^{2} + \dot{\phi}_{1}(t,s)^{2}\right) \notag \\
	&= \sum_{i,j,k=0}^{\infty} \ddot{c}_{k}(t) \int_{0}^{1} \diff{r} r e_{l}(r) e_{k}(r) \notag \\
	&\quad \int_{0}^{r} \diff{s} s^{2} \left[ c_{i}(t) c_{j}(t) e_{i}'(s) e_{j}'(s)  + \dot{c}_{i}(t) \dot{c}_{j}(t) e_{i}(s) e_{j}(s)\right]\,,
\end{align}

\begin{align}
	\label{eq:20.12.18_36}
	\inner{\delta_{2}\ddot{\phi}_{1}}{e_{l}} &= -\int_{0}^{1} \diff{r} r^{2} e_{l}(r) \ddot{\phi}_{1}(t,r) \int_{0}^{r} \diff{s} s \left(\phi_{1}'(t,s)^{2} + \dot{\phi}_{1}(t,s)^{2}\right) \notag \\
	&= -\sum_{i,j,k=0}^{\infty} \ddot{c}_{k}(t) \int_{0}^{1} \diff{r} r^{2} e_{l}(r) e_{k}(r) \notag \\
	&\quad \int_{0}^{r} \diff{s} s \left[ c_{i}(t) c_{j}(t) e_{i}'(s) e_{j}'(s)  + \dot{c}_{i}(t) \dot{c}_{j}(t) e_{i}(s) e_{j}(s)\right]\,.
\end{align}

Next, we define different types of integrals appearing in \eqref{eq:20.12.18_32}-\eqref{eq:20.12.18_36} as
\begin{align}
	\label{eq:20.12.20_01}
	M^{*}_{klij} &=\int_{0}^{1}\diff{r}e_{l}e_{k}'\int_{0}^{r}\diff{s}s^{2}e_{i}'e_{j}'\,,
	\\
	\label{eq:20.12.20_02}
	M_{klij} &= \int_{0}^{1}\diff{r}e_{l}e_{k}'\int_{0}^{r}\diff{s}s^{2}e_{i}e_{j}\,,
	\\
	\label{eq:20.12.20_03}
	K^{*}_{klij} &= \int_{0}^{1}\diff{r}re_{l}e_{k}\int_{0}^{r}\diff{s}s^{2}e_{i}'e_{j}'\,,
	\\
	\label{eq:20.12.20_04}
	K_{klij} &= \int_{0}^{1}\diff{r}re_{l}e_{k}\int_{0}^{r}\diff{s}s^{2}e_{i}e_{j}\,,
	\\
	\label{eq:20.12.20_05}
	L^{*}_{klij} &= \int_{0}^{1}\diff{r}r^{2}e_{k}e_{l}\int_{0}^{r}\diff{s}se_{i}'e_{j}'\,,
	\\
	\label{eq:20.12.20_06}
	L_{klij} &= \int_{0}^{1}\diff{r}r^{2}e_{k}e_{l}\int_{0}^{r}\diff{s}se_{i}e_{j}\,.
\end{align}
(Note that here and in the following expressions, the asterisk $^{*}$ does not indicate complex conjugation, all the integrals are real.) Using this, we find
\begin{multline}
	\label{eq:20.12.30_01}
	\inner{S_{3}}{e_{l}} = - \sum_{ijk}\left[
	- c_{k}c_{i}c_{j}M^{*}_{klij} - c_{k}\dot{c}_{i}\dot{c}_{j}M_{klij}
	\right.
	\\
	+ \dot{c}_{k}(c_{i}c_{j})\dot{}K^{*}_{klij}
	+ \dot{c}_{k}(\dot{c}_{i}\dot{c}_{j})\dot{}\,K_{klij} 
	\\
	+ \left(-\dot{c}_{k}(c_{i}c_{j})\dot{}\,L^{*}_{klij}-\dot{c}_{k}(\dot{c}_{i}\dot{c}_{j})\dot{}\,L_{klij}\right)
	\\
	\left.
	+ 2\left(\ddot{c}_{k}c_{i}c_{j}K^{*}_{klij} + \ddot{c}_{k}\dot{c}_{i}\dot{c}_{j}K_{klij} - \ddot{c}_{k}c_{i}c_{j}L^{*}_{klij} - \ddot{c}_{k}\dot{c}_{i}\dot{c}_{j}L_{klij}\right)
	\right]\,.
\end{multline}
To simplify the source term \eqref{eq:20.12.30_01}, we make use of the identities
\begin{align}
	\label{eq:20.12.20_07}
	\dot{c}_{k} \left(c_{i}c_{j}\right)\dot{}  &= \dot{c}_{i}c_{j}\dot{c}_{k} + c_{i}\dot{c}_{j}\dot{c}_{k}\,,
	\\
	\label{eq:20.12.20_08}
	\dot{c}_{k} \left(\dot{c}_{i}\dot{c}_{j}\right)\dot{}  &= -\omega_{i}^{2}c_{i}\dot{c}_{j}\dot{c}_{k} - \omega_{j}^{2} \dot{c}_{i}c_{j}\dot{c}_{k}\,,
	\\
	\label{eq:20.12.20_09}
	c_{i}c_{j}\ddot{c}_{k} &= - \omega_{k}^{2}c_{i}c_{j}c_{k}\,,
	\\
	\label{eq:20.12.20_10}
	\dot{c}_{i}\dot{c}_{j}\ddot{c}_{k} &= - \omega_{k}^{2}\dot{c}_{i}\dot{c}_{j}c_{k}\,,
\end{align}
and the definition \eqref{eq:21.08.02_01} and write
\begin{multline}
	\label{eq:21.08.02_02}
	\inner{S_{3}}{e_{l}} = \left(e^{i(\omega_{i}-\omega_{j}-\omega_{k})t}\alpha_{i}\bar{\alpha}_{j}\bar{\alpha}_{k} + c.c.\right)O_{klij}
	+ \left(e^{i(\omega_{i}+\omega_{j}-\omega_{k})t}\alpha_{i}\alpha_{j}\bar{\alpha}_{k} + c.c.\right)P_{klij}
	\\
	+ \left(e^{i(\omega_{i}-\omega_{j}+\omega_{k})t}\alpha_{i}\bar{\alpha}_{j}\alpha_{k} + c.c.\right)Q_{klij}
	+
	\left(e^{i(\omega_{i}+\omega_{j}+\omega_{k})t}\alpha_{i}\alpha_{j}\alpha_{k} + c.c.\right)R_{klij}
	\,,
\end{multline}
where we defined
\begin{multline}
	\label{eq:21.08.02_03}
	O_{klij} = \omega _i \omega _j \omega _k \left(-\omega _i+\omega _j+2
   \omega _k\right) K_{klij}+\omega _k \left(-\omega
   _i+\omega _j+2 \omega _k\right) K^{*}_{klij}
   \\
   +\omega
   _i \omega _j \omega _k \left(\omega _i-\omega _j-2 \omega
   _k\right) L_{klij}+\omega _k \left(\omega _i-\omega _j-2
   \omega _k\right) L^{*}_{klij}
   \\
   +\omega _i \omega _j M_{klij}+M^{*}_{klij}\,,
\end{multline}

\begin{multline}
	\label{eq:21.08.02_04}
   P_{klij} = \omega _i \omega _j \omega _k \left(\omega _i+\omega _j-2
   \omega _k\right) K_{klij}+\omega _k \left(-\omega
   _i-\omega _j+2 \omega _k\right) K^{*}_{klij}
   \\
   -\omega
   _i \omega _j \omega _k \left(\omega _i+\omega _j-2 \omega
   _k\right) L_{klij}+\omega _k \left(\omega _i+\omega _j-2
   \omega _k\right) L^{*}_{klij}
   \\
   -\omega _i \omega _j
   M_{klij}+M^{*}_{klij}\,,
\end{multline}

\begin{multline}
   \label{eq:21.08.02_05}
   Q_{klij} = \omega _i \omega _j \omega _k \left(\omega _i-\omega _j+2
   \omega _k\right) K_{klij}+\omega _k \left(\omega
   _i-\omega _j+2 \omega _k\right) K^{*}_{klij}
   \\
   +\omega
   _i \omega _j \omega _k \left(-\omega _i+\omega _j-2 \omega
   _k\right) L_{klij}+\omega _k \left(-\omega _i+\omega _j-2
   \omega _k\right) L^{*}_{klij}
   \\
   +\omega _i \omega _j
   M_{klij}+M^{*}_{klij}\,,
\end{multline}

\begin{multline}
   \label{eq:21.08.02_06}
   R_{klij} = -\omega _i \omega _j \omega _k \left(\omega _i+\omega _j+2
   \omega _k\right) K_{klij}+\omega _k \left(\omega
   _i+\omega _j+2 \omega _k\right) K^{*}_{klij}
   \\
   +\omega
   _i \omega _j \omega _k \left(\omega _i+\omega _j+2 \omega
   _k\right) L_{klij}-\omega _k \left(\omega _i+\omega _j+2
   \omega _k\right) L^{*}_{klij}
   \\
   -\omega _i \omega _j
   M_{klij}+M^{*}_{klij}\,.
\end{multline}
Renaming the dummy indices $i$ and $j$ in the third and sixth term, and $i$ and $k$ in the fourth and seventh term, we rewrite \eqref{eq:21.08.02_02} as
\begin{align}
	\label{eq:21.01.07_07}
	\inner{S_{3}}{e_{l}} &= \sum_{ijk} \left[\alpha_{i} \alpha_{j} \alpha_{k} e^{i(\omega_{i}+\omega_{j}+\omega_{k})t} R_{klij} \right. \notag \\
	&\qquad \; \left. +\bar{\alpha}_{i} \alpha_{j} \alpha_{k} e^{i(-\omega_{i}+\omega_{j}+\omega_{k})t} \left(O_{klij} + Q_{klji} + P_{ilkj} \right) \right. \notag \\
	&\qquad \; \left. + \alpha_{i} \bar{\alpha}_{j} \bar{\alpha}_{k} e^{i(\omega_{i}-\omega_{j}-\omega_{k})t} \left(O_{klij} + Q_{klji} + P_{ilkj} \right) \right. \notag \\
	&\qquad \; \left. + \bar{\alpha}_{i} \bar{\alpha}_{j} \bar{\alpha}_{k} e^{i(-\omega_{i}-\omega_{j}-\omega_{k})t} R_{klij} \right] \\
	\label{eq:21.01.08_01}
	&= \sum_{ijk} \left[\alpha_{i} \alpha_{j} \alpha_{k} e^{i(\omega_{i}+\omega_{j}+\omega_{k})t} R_{klij} \right. \notag \\
	&\qquad \; \left. +\bar{\alpha}_{i} \alpha_{j} \alpha_{k} e^{i(-\omega_{i}+\omega_{j}+\omega_{k})t} S_{klij} \right. \notag \\
	&\qquad \; \left. + \alpha_{i} \bar{\alpha}_{j} \bar{\alpha}_{k} e^{i(\omega_{i}-\omega_{j}-\omega_{k})t} S_{klij} \right. \notag \\
	&\qquad \; \left. + \bar{\alpha}_{i} \bar{\alpha}_{j} \bar{\alpha}_{k} e^{i(-\omega_{i}-\omega_{j}-\omega_{k})t} R_{klij} \right]\,,
\end{align}
where we introduced 
\begin{equation}
	\label{eq:21.08.02_07}
	S_{klij} := O_{klij} + Q_{klji} + P_{ilkj}
	\,.
\end{equation}

Now observe that whenever a combination of eigenfrequencies satisfies $\pm \omega_{i} \pm \omega_{j}\pm\omega_{k} = \pm\omega_{l}$ and the respective coefficient in \eqref{eq:21.01.07_07} does not vanish, the term is in resonance with the mode $e_{l}$. This then produces a secular term in the solution $c^{(3)}_{l}(t)\sim t$ and spoils the perturbative expansion \eqref{eq:20.12.18_08}-\eqref{eq:20.12.18_10}.

We now use multiple-scale analysis \cite{Murdock.1999book} to derive the resonant system. We introduce the `slow time' dependence $\tau = \varepsilon^{2} t$ and write
\begin{equation}
	\label{eq:21.01.07_01}
	\phi = \phi(t,\tau,r)\,,
\end{equation}
similarly for $A$, $\delta$. We treat $\tau$ as an independent variable, so in particular
\begin{equation}
	\label{eq:21.01.07_03}
	\partial_{t}^{2} \phi(t,r) \rightarrow \partial_{t}^{2} \phi(t,\tau,r) + 2 \varepsilon^{2} \partial_{t} \partial_{\tau} \phi(t,\tau,r) + \varepsilon^{4} \partial_{\tau}^{2} \phi(t,\tau,r)
	\,.
\end{equation}
Expanding $\phi$, $A$, $\delta$ as above, we get the same equations as before at orders $\varepsilon$ and $\varepsilon^{2}$. However, the solution to the homogeneous equation at the first order of $\varepsilon$ is now
\begin{equation}
	\label{eq:21.01.07_02}
	\phi_{1}(t,\tau,r) = \sum_{n \geq 0} \left(\alpha_{n}(\tau) e^{i\omega_{n}t} + \bar{\alpha}_{n}(\tau) e^{-i\omega_{n}t} \right) e_{n}(r) \equiv \sum_{n \geq 0} c_{n}(t,\tau) e_{n}(r)
	\,,
\end{equation}
cf. \eqref{eq:20.12.18_19}. At the third order, using \eqref{eq:21.01.07_03}, we get the equation
\begin{equation}
	\label{eq:21.01.07_04}
	\ddot{\phi}_{3} + L \phi_{3} = -2 \partial_{t} \partial_{\tau} \phi_{1} - (A_{2}'+\delta_{2}')\phi_{1}' 
	- (\dot{A}_{2}+\dot{\delta}_{2})\dot{\phi}_{1} 
	- 2(A_{2}+\delta_{2})\ddot{\phi}_{1}\,,
\end{equation}
which now contains a new term, cf. \eqref{eq:20.12.18_15}. We project this equation onto the eigenbasis $\{e_{l} \}$ and get the system of equations
\begin{equation}
	\label{eq:21.01.07_05}
	\ddot{c}_{l}^{(3)} + \omega_{l}^{2} c_{l}^{(3)} = \inner{-2 \partial_{t} \partial_{\tau} \phi_{1} + S_{3}}{e_l}\,.
\end{equation}
To remove the resonant terms, we set the projection of \eqref{eq:21.01.07_05} onto the Fourier mode $e^{i \omega_{l}t}$ to zero. This yields
\begin{equation}
	\label{eq:21.01.08_03}
	\int_{0}^{2} \diff{t} \inner{-2 \partial_{t} \partial_{\tau} \phi_{1} + S_{3}}{e_l} e^{-i \omega_{l}t} = 0\,,
\end{equation}
which represents a condition on the coefficients $\alpha_{j}(\tau)$ that, if fulfilled, eliminates secular terms from the solution to \eqref{eq:21.01.07_05}. For the first term in \eqref{eq:21.01.08_03} we have
\begin{align}
	\label{eq:21.01.08_04}
	\int_{0}^{2} \diff{t} \inner{-2 \partial_{t} \partial_{\tau} \phi_{1}}{e_l} e^{-i \omega_{l}t} &= - 2i \omega_{l} \int_{0}^{2} \diff{t} \left(\frac{\diff}{\diff{\tau}}\alpha_{l}(\tau) e^{i \omega_{l}t} - \frac{\diff}{\diff{\tau}}\bar{\alpha}_{l}(\tau) e^{-i \omega_{l}t} \right) e^{-i \omega_{l}t} \notag \\
	&= - 4 i \omega_{l} \frac{\diff}{\diff{\tau}}\alpha_{l}(\tau)
	\,.
\end{align}
For the second term in \eqref{eq:21.01.08_03}, we get
\begin{align}
	\label{eq:21.01.08_05}
	\int_{0}^{2} \diff{t} \inner{S_{3}}{e_l} e^{-i \omega_{l}t} = 2 \sum_{ijk} &\left[\alpha_{i} \alpha_{j} \alpha_{k} R_{klij} \delta_{l+1,i+1+j+1+k+1} \right. \notag \\
	&\left. \! \! \! \! +\bar{\alpha}_{i} \alpha_{j} \alpha_{k} S_{klij} \delta_{l+1,-(i+1)+j+1+k+1} \right. \notag \\
	&\left. \! \! \! \! + \alpha_{i} \bar{\alpha}_{j} \bar{\alpha}_{k} S_{klij} \delta_{l+1,i+1-(j+1)-(k+1)} \right. \notag \\
	&\left. \! \! \! \! + \bar{\alpha}_{i} \bar{\alpha}_{j} \bar{\alpha}_{k} R_{klij} \delta_{l+1,-(i+1)-(j+1)-(k+1)} \right]\,,
\end{align}
where we notice that $\delta_{l+1,-(i+1)-(j+1)-(k+1)} = 0$ $\forall \, i,j,k,l =0,1,\ldots$. Putting \eqref{eq:21.01.08_04} and \eqref{eq:21.01.08_05} together, \eqref{eq:21.01.08_03} becomes
\begin{equation}
	\label{eq:21.01.08_06}
	2i\omega_{l}\frac{\diff{} \alpha_{l}}{\diff{\tau}} =\sum_{ijk}^{+++} R_{klij} \alpha_{i} \alpha_{j} \alpha_{k} + \sum_{ijk}^{-++} S_{klij} \bar{\alpha}_{i} \alpha_{j} \alpha_{k} + \sum_{ijk}^{+--} S_{klij} \alpha_{i} \bar{\alpha}_{j} \bar{\alpha}_{k}\,,
\end{equation}
where we use the following notation for resonant sums:
\begin{equation}
	\label{eq:21.01.08_07}
	\sum_{ijk}^{+++} = \sum_{\substack{i,j,k=0 \\ l+1=i+1+j+1+k+1}}^{\infty}\,, 
	\quad
	\sum_{ijk}^{-++} = \sum_{\substack{i,j,k=0 \\ l+1=-(i+1)+j+1+k+1}}^{\infty}\,,
	\quad
	\sum_{ijk}^{+--} = \sum_{\substack{i,j,k=0 \\ l+1=i+1-(j+1)-(k+1)}}^{\infty}\,.
\end{equation}
Equation \eqref{eq:21.01.08_06} is equivalent to the renormalization flow equations as derived in \cite{Craps.2014, Craps.2015}.
To bring \eqref{eq:21.01.08_06} to a canonical form we define 
\begin{equation}
	\label{eq:21.08.02_08}
	S^{+++}_{ijkl} = R_{klij}\,,
	\quad
	S^{++-}_{ijkl} = S_{ilkj}\,,
	\quad
	S^{+--}_{ijkl} = S_{klij}
	\,.
\end{equation}
Then we can write \eqref{eq:21.01.08_06} as
\begin{equation}
	\label{eq:21.08.02_09}
	2i\omega_{l}\frac{\diff{} \alpha_{l}}{\diff{\tau}} = 
	\sum_{ijk}^{+++} S^{+++}_{ijkl} \alpha_{i} \alpha_{j} \alpha_{k} +
	\sum_{ijk}^{++-} S^{++-}_{ijkl} \alpha_{i} \alpha_{j} \bar{\alpha}_{k} +
	\sum_{ijk}^{+--} S^{+--}_{ijkl} \alpha_{i} \bar{\alpha}_{j} \bar{\alpha}_{k}
	\,.
\end{equation}

In parallel to studying \eqref{eq:21.08.02_09}, which we refer to as the full resonant system, we study the $++-$ resonant system, where we drop the $+++$ and $+--$ terms from \eqref{eq:21.08.02_09}. Thus we consider
\begin{equation}
	\label{eq:22.03.15_01}
	2i\omega_{l}\frac{\diff{} \alpha_{l}}{\diff{\tau}} = 	
	\sum_{ijk}^{++-} S^{++-}_{ijkl} \alpha_{i} \alpha_{j} \bar{\alpha}_{k}
	\,.
\end{equation}
Below, we will argue that \eqref{eq:22.03.15_01} is a good approximation to \eqref{eq:21.08.02_09} whenever the solution develops a singularity.

Note that both the system \eqref{eq:21.08.02_09} and \eqref{eq:22.03.15_01} are invariant under
\begin{equation}
	\label{eq:22.07.20_01}
	\alpha_{l}(\tau) \rightarrow \varepsilon^{-1}\alpha_{l}(\varepsilon^{-2}\tau)
	\,,
\end{equation}
thus a single solution allows us to conclude about the behavior of solutions in the limit $\varepsilon\rightarrow 0$. A comparison with the solution of the Einstein-Scalar field system suggests that the resonant approach provides a good approximation on the timescale $\mathcal{O}(\varepsilon^{-2})$. The details are presented in Sec.~\ref{sec:Results}.

It is convenient to rewrite the $++-$ sum, and in particular the system \eqref{eq:22.03.15_01}, as \cite{Craps.2015}
\begin{equation}
	\label{eq:22.03.15_02}
	2i\omega_{l}\frac{\diff{} \alpha_{l}}{\diff{\tau}} = T_{l}|\alpha_{l}|^{2}\alpha_{l} + \sum_{i\neq l}R_{il}|\alpha_{i}|^{2}\alpha_{l} + \sum_{ijk}^{++-}{}^{'} S^{++-}_{ijkl} \alpha_{i} \alpha_{j} \bar{\alpha}_{k}\,,
\end{equation}
where
\begin{equation}
	\label{eq:22.03.15_03}
	T_{l} = S^{++-}_{llll}\,, \quad R_{il} = S^{++-}_{ilil} + S^{++-}_{liil}
	\,,
\end{equation}
and the \textit{primed sum} is
\begin{equation}
	\label{eq:22.03.15_04}
	\sum_{ijk}^{++-}{}^{'} = \underbrace{\sum_{ijkl}}_{\substack{i+j-k=l\\ i\neq k \wedge i\neq l}}
	\,.
\end{equation}

\subsection{Interaction coefficients}
\label{sec:InteractionCoefficients}
Evaluating the integrals \eqref{eq:20.12.20_01}-\eqref{eq:20.12.20_06} (some of which need to be computed for several special cases) and using the definitions \eqref{eq:21.08.02_03}-\eqref{eq:21.08.02_06} and \eqref{eq:21.08.02_07}, we obtain an explicit form of the interaction coefficients.
For the resonant combination $++-$ we get
\begin{equation}
	\label{eq:22.06.06_01}
	T_{i} = \omega _i^4 \left(8 \text{Ci}\left(2 \omega _i\right)-8 \log \left(2 \omega _i\right)-8 \gamma
   +20\right)+10 \omega _i^3 \left(\text{Si}\left(4 \omega _i\right)-2 \text{Si}\left(2 \omega
   _i\right)\right)
   \,,
\end{equation}
\begin{multline}
	\label{eq:22.06.06_02}
	R_{il} = 
	-4 \omega _i^2 \omega _l^2 \left(-2 \text{Ci}\left(2 \omega _i\right)+2 \log \left(\omega
   _i\right)+2 \gamma -5+\log (4)\right)
   \\
   +\frac{8 \omega _l \left(2 \omega _l^4-\left(\omega
   _i^2-\omega _l^2\right){}^2\right) \text{Si}\left(2 \omega _l\right)}{\omega _i^2-\omega
   _l^2}
   -\frac{8 \omega _i \left(2 \omega _i^4-\left(\omega _i^2-\omega _l^2\right){}^2\right)
   \text{Si}\left(2 \omega _i\right)}{\omega _i^2-\omega _l^2}
   \\
   +\frac{2 \left(\left(\omega
   _i-\omega _l\right){}^4+\left(\omega _i^2+\omega _l^2\right){}^2\right) \text{Si}\left(2
   \left(\omega _i-\omega _l\right)\right)}{\omega _i-\omega _l}
   \\
   +\frac{2 \left(\left(\omega
   _i+\omega _l\right){}^4+\left(\omega _i^2+\omega _l^2\right){}^2\right) \text{Si}\left(2
   \left(\omega _i+\omega _l\right)\right)}{\omega _i+\omega _l}
	\,,
\end{multline}
recall \eqref{eq:22.03.15_03}, and for $k=i+j-l$ with $i\neq l \wedge j\neq l$ we find
\begin{multline}
	\label{eq:22.06.06_03}
S^{++-}_{ijkl} = 
\frac{2 \left(\omega _i^2+\omega _l^2\right) \left(-2 \omega _l \left(\omega _i+\omega
   _j\right)+2 \omega _i \omega _j+\omega _i^2+2 \omega _j^2+\omega _l^2\right)
   \text{Si}\left(2 \left(\omega _i-\omega _l\right)\right)}{\omega _i-\omega
   _l}
   \\
   +\left(\frac{2 \omega _l^2 \left(\omega _i^2+\omega _j^2\right)}{\omega _i+\omega _j}-2
   \omega _l \left(\omega _i^2+\omega _j^2\right)+2 \left(\omega _i+\omega _j\right)
   \left(\omega _i \omega _j+\omega _i^2+\omega _j^2\right)\right) \text{Si}\left(2
   \left(\omega _i+\omega _j\right)\right)
   \\
   +\text{Si}\left(2 \omega _l\right) \left(\frac{2
   \omega _j \omega _l^2 \left(3 \omega _i+2 \omega _j\right)}{\omega _i+\omega _j}+\frac{4
   \omega _i^2 \omega _j^2}{\omega _i-\omega _l}-4 \omega _i \omega _j^2-4 \omega _j^2 \omega
   _l-2 \omega _l^3\right)
   \\
   +2 \omega _j^2 \text{Si}\left(2 \omega _j\right) \left(\omega _l
   \left(-\frac{\omega _l}{\omega _i+\omega _j}+\frac{2 \omega _i}{\omega _i-\omega
   _l}+1\right)-\omega _j\right)
   \\
   -\frac{2 \left(\omega _i+\omega _j-\omega _l\right){}^2
   \left(\omega _i \left(\omega _j^2+\omega _l^2\right)+\omega _i^2 \omega _j+\omega
   _i^3+\omega _j \omega _l \left(\omega _l-\omega _j\right)\right) \text{Si}\left(2
   \left(\omega _i+\omega _j-\omega _l\right)\right)}{\left(\omega _i+\omega _j\right)
   \left(\omega _i-\omega _l\right)}
   \\
   +2 \omega _i^2 \text{Si}\left(2 \omega _i\right)
   \left(-\frac{\omega _l^2}{\omega _i+\omega _j}+2 \omega _j \left(\frac{\omega _j}{\omega
   _l-\omega _i}-1\right)-\omega _i+\omega _l\right)
   \\
   +2 \left(\omega _j-\omega _l\right){}^3
   \text{Si}\left(2 \left(\omega _j-\omega _l\right)\right)
   \,.
\end{multline}

For the resonant combination $k=i-j-l-2$, we get
\begin{multline}
	\label{eq:22.06.06_04}
	S^{+--}_{ijkl} = -\frac{2 \left(\omega _i^2+\omega _j^2\right) \left(2 \omega _l \left(\omega _j-\omega
   _i\right)+\left(\omega _i-\omega _j\right){}^2+2 \omega _l^2\right) \text{Si}\left(2
   \omega _i-2 \omega _j\right)}{\omega _i-\omega _j}
   \\
   +2 \text{Si}\left(2 \omega
   _l\right) \left(\frac{\omega _j \omega _l^2 \left(3 \omega _i-\omega
   _j\right)}{\omega _i-\omega _j}+\frac{\omega _i^2 \omega _j^2}{\omega _l-\omega
   _i}+\omega _i \omega _j^2+\omega _j^2 \omega _l+\omega _l^3\right)
   \\
   +2 \omega _i^2
   \text{Si}\left(2 \omega _i\right) \left(\omega _j \left(\frac{\omega _j}{\omega
   _i-\omega _l}-1\right)+2 \omega _l \left(\frac{\omega _l}{\omega _i-\omega
   _j}-1\right)+\omega _i\right)
   \\
   +2 \omega _j^2 \text{Si}\left(2 \omega _j\right)
   \left(\omega _l \left(\frac{2 \omega _l}{\omega _j-\omega _i}+\frac{\omega _i}{\omega
   _i-\omega _l}+2\right)+\omega _j\right)
   \\
   +\frac{2 \left(-\omega _i+\omega _j+\omega
   _l\right){}^2 \left(\omega _i \left(\omega _j^2+\omega _l^2\right)-\omega _i^2 \omega
   _l+\omega _i^3-\omega _j \omega _l \left(\omega _j+\omega _l\right)\right)
   \text{Si}\left(2 \omega _i-2 \omega _j-2 \omega _l\right)}{\left(\omega _i-\omega
   _j\right) \left(\omega _i-\omega _l\right)}
   \\
   +2 \left(\omega _i^2 \left(\frac{2 \omega
   _j^2}{\omega _l-\omega _i}+\omega _j+2 \omega _l\right)+\omega _i \left(\omega _j^2-2
   \omega _l^2\right)-\omega _i^3+\omega _l \left(\omega _j \omega _l+\omega _j^2+\omega
   _l^2\right)\right) \text{Si}\left(2 \omega _i-2 \omega _l\right)
   \\
   -2 \left(\omega
   _j+\omega _l\right){}^3 \text{Si}\left(2 \left(\omega _j+\omega _l\right)\right)
   \,,
\end{multline}
(note that both $i=j$ and $i=l$ are excluded by the $+--$ resonant condition), whereas for $k=l-i-j-2$ we have
\begin{multline}
	\label{eq:22.06.06_05}
	S^{+++}_{ijkl} = 
	\frac{2}{3} \omega _l^2 \text{Si}\left(2 \omega _l\right) \left(\omega _l-\frac{3 \omega
   _i \omega _j}{\omega _i+\omega _j}\right)
   \\
   +\left(-\frac{2 \omega _l^2 \left(\omega
   _i^2+\omega _j^2\right)}{\omega _i+\omega _j}+2 \omega _l \left(\omega _i^2+\omega
   _j^2\right)-\frac{2}{3} \left(\omega _i^3+\omega _j^3\right)\right) \text{Si}\left(2
   \left(\omega _i+\omega _j\right)\right)
   \\
   +\frac{2 \left(\omega _i+\omega _j-\omega
   _l\right){}^2 \left(-\omega _i \left(\omega _j+\omega _l\right)+\omega _i^2+\omega _j
   \left(\omega _j-\omega _l\right)\right) \text{Si}\left(2 \left(\omega _i+\omega
   _j-\omega _l\right)\right)}{3 \left(\omega _i+\omega _j\right)}
   \\
   +\frac{2}{3} \omega
   _i^2 \text{Si}\left(2 \omega _i\right) \left(3 \omega _l \left(\frac{\omega
   _l}{\omega _i+\omega _j}-1\right)+\omega _i\right)+\frac{2}{3} \omega _j^2
   \text{Si}\left(2 \omega _j\right) \left(3 \omega _l \left(\frac{\omega _l}{\omega
   _i+\omega _j}-1\right)+\omega _j\right)
   \\
   -\frac{2}{3} \left(\omega _i-\omega
   _l\right){}^3 \text{Si}\left(2 \omega _i-2 \omega _l\right)-\frac{2}{3} \left(\omega
   _j-\omega _l\right){}^3 \text{Si}\left(2 \omega _j-2 \omega _l\right)
   \,.
\end{multline}
The functions $\Si$ and $\Ci$ appearing in \eqref{eq:22.06.06_01}-\eqref{eq:22.06.06_05} are the trigonometric integrals \cite[\href{https://dlmf.nist.gov/6.2\#ii}{Sec.~6.2(ii)}]{NIST:DLMF} 
\begin{equation}
	\label{eq:22.07.23_03}
	\Si(x) = \int_{0}^{x}\frac{\sin{y}}{y}\diff{y}\,,
	\quad
	\Ci(x) = -\int_{x}^{\infty}\frac{\cos{y}}{y}\diff{y}\,.
\end{equation}
and $\gamma$ is the Euler-Mascheroni constant \cite[\href{https://dlmf.nist.gov/5.2.ii}{Sec.~5.2(ii)}]{NIST:DLMF}.

The explicit evaluation of \eqref{eq:22.06.06_04}-\eqref{eq:22.06.06_05} shows that the $+--$ and $+++$ terms are generically non-zero, as opposed to the AdS case \cite{Craps.2014, Craps.2015}, where they vanish as a result of the high symmetry of the background geometry.
As a consequence, the additional terms in the equation \eqref{eq:21.08.02_09} in combination with the specific symmetry of interaction coefficients with respect to a permutation of indices, influence the set of (known) integrals of motion of the resonant system \eqref{eq:21.08.02_09} and \eqref{eq:22.03.15_01}. This is different from the asymptotically AdS case, where the additional terms are absent and the coefficients have a different symmetry \cite{Evnin.2021v}.

\subsection{Conserved quantities}
\label{sec:ConservedQuantities}

We split the discussion of conserved quantities into two types of resonant systems: the full resonant system \eqref{eq:21.08.02_09} and the $++-$ system \eqref{eq:22.03.15_01}.

\subsubsection{The $++-$ resonant system}
In the following, we prove that $E=\sum_{i}\omega_{i}^{2}|\alpha_{i}|^2$ is constant along the flow generated by \eqref{eq:22.03.15_01}. Using the product rule and inserting \eqref{eq:22.03.15_02}, we find
\begin{multline}
	\label{eq:22.03.15_05}
	\frac{\diff{}}{\diff{\tau}}E = \sum_{l}\omega_{l}^{2}\left(\dot{\alpha}_{l}\bar{\alpha}_{l}+\alpha_{l}\dot{\bar{\alpha}}_{l}\right)
	= \frac{1}{2i}\sum_{l}\omega_{l}\sum_{ijk}^{++-}{}^{'}S^{++-}_{ijkl}\left(\alpha_{i}\alpha_{j}\bar{\alpha}_{k}\bar{\alpha}_{l} - \bar{\alpha}_{i}\bar{\alpha}_{j}\alpha_{k}\alpha_{l}\right)
	\\
	= \frac{1}{4i}\sum_{l}\sum_{ijk}^{++-}{}^{'}\left(S^{++-}_{ijkl}\omega_{l} + S^{++-}_{ijlk}\omega_{k}\right)\left(\alpha_{i}\alpha_{j}\bar{\alpha}_{k}\bar{\alpha}_{l} - \bar{\alpha}_{i}\bar{\alpha}_{j}\alpha_{k}\alpha_{l}\right)
	\,,
\end{multline}
where in the second step, we note that the $T_{l}$ and $R_{il}$ terms cancel out, leaving only the primed sum \eqref{eq:22.03.15_04}. In the last step, we used the symmetry of the $\alpha$ term with respect to the indices $k$ and $l$. Next, using the identity
\begin{equation}
	\label{eq:22.03.16_06}
	S^{++-}_{ijkl} = S^{++-}_{jilk}, \quad i+j-k=l\,, \ i\neq l\,,\ j\neq l\,,
\end{equation}
we rewrite the sum as
\begin{multline}
	\label{eq:22.03.15_07}
	\frac{\diff{}}{\diff{\tau}}E = \frac{1}{4i}\sum_{l}\sum_{ijk}^{++-}{}^{'}\left(S^{++-}_{ijkl}\omega_{l} + S^{++-}_{jikl}\omega_{k}\right)\left(\alpha_{i}\alpha_{j}\bar{\alpha}_{k}\bar{\alpha}_{l} - \bar{\alpha}_{i}\bar{\alpha}_{j}\alpha_{k}\alpha_{l}\right)
	\\
	= \frac{1}{4i}\sum_{l}\sum_{ijk}^{++-}{}^{'}S^{++-}_{ijkl}\left(\omega_{l} + \omega_{k}\right)\left(\alpha_{i}\alpha_{j}\bar{\alpha}_{k}\bar{\alpha}_{l} - \bar{\alpha}_{i}\bar{\alpha}_{j}\alpha_{k}\alpha_{l}\right)\,,
\end{multline}
where in the last step, we used the permutation of indices $i\leftrightarrow j$ in $S^{++-}_{jikl}$ and the symmetry of the $\alpha$ term to factorize the $S^{++-}_{ijkl}$. In the next step, we use another identity satisfied by the interaction coefficients, namely
\begin{equation}
	\label{eq:22.03.15_08}
	S^{++-}_{ijkl} = S^{++-}_{klij}, \quad i+j-k=l\,, \ i\neq l\,,\ j\neq l
	\,.
\end{equation}
It follows that the product $S^{++-}_{ijkl}\left(\omega_{l} + \omega_{k}\right)$ is symmetric under the pair interchange $(i,j)\leftrightarrow(k,l)$, when restricted to the resonant combination under the primed sum. Since the $\alpha$ term is antisymmetric under this transformation, the contraction in \eqref{eq:22.03.15_07} vanishes, and we have
\begin{equation}
	\label{eq:22.03.15_09}
	\frac{\diff{}}{\diff{\tau}}E = 0\,.
\end{equation}

In a similar way, we show that $J=\sum_{i}\omega_{i}|\alpha_{i}|^{2}$ is conserved. We have
\begin{equation}
	\label{eq:22.03.16_01}
	\frac{\diff{}}{\diff{\tau}}J = \sum_{l}\left(\omega_{l}\dot{\alpha}_{l}\bar{\alpha}_{l}+ \omega_{l}\alpha_{l}\dot{\bar{\alpha}}_{l}\right) = \frac{1}{2i}\sum_{l}\sum_{ijk}^{++-}{}^{'}S^{++-}_{ijkl}\left(\alpha_{i}\alpha_{j}\bar{\alpha}_{k}\bar{\alpha}_{l}-\bar{\alpha}_{i}\bar{\alpha}_{j}\alpha_{k}\alpha_{l}\right)\,,
\end{equation}
since the $T_{l}$ and $R_{il}$ terms cancel and the remaining sum is \eqref{eq:22.03.15_04}, as in the proof of $\dot{E}=0$. Because of the symmetry \eqref{eq:22.03.15_08} of the coefficients $S_{ijkl}^{++-}$ under the pair interchange $(i,j)\leftrightarrow(k,l)$ for the resonant condition $i+j-k=l$ with $i\neq k \wedge i\neq l$, the contraction in \eqref{eq:22.03.16_01} vanishes, so
\begin{equation}
	\label{eq:22.03.16_02}
	\frac{\diff{}}{\diff{\tau}}J = 0\,.
\end{equation}

Next, we look for another conserved quantity. First, we define
\begin{multline}
	\label{eq:22.03.16_03}
	V = \sum_{l}\sum^{++-}_{ijk} S^{++-}_{ijkl} \alpha_{i}\alpha_{j}\bar{\alpha}_{k}\bar{\alpha}_{l}
	\\ 
	= \sum_{l} T_{l}|\alpha_{l}|^{2}|\alpha_{l}|^{2} + \sum_{l}\sum_{i\neq l}R_{il}|\alpha_{i}|^{2}|\alpha_{l}|^{2} + \sum_{l}\sum_{ijk}^{++-}{}^{'} S^{++-}_{ijkl} \alpha_{i} \alpha_{j} \bar{\alpha}_{k}\bar{\alpha}_{l}
	\,.
\end{multline}
Taking the derivative of $V$ with respect to $\bar{\alpha}_{n}$, we find
\begin{multline}
	\label{eq:22.03.16_04}
	\frac{\partial V}{\partial \bar{\alpha}_{n}} = 2T_{n}|\alpha_{n}|^{2}\alpha_{n} + \sum_{i\neq n}\left(R_{in}+R_{ni}\right)|\alpha_{i}|^{2}\alpha_{n} + \sum^{++-}_{ijk}{}^{'} \left(S^{++-}_{ijkn} + S^{++-}_{ijnk}\right)\alpha_{i}\alpha_{j}\bar{\alpha}_{k}
	\\
	= 2T_{n}|\alpha_{n}|^{2}\alpha_{n} + \sum_{i\neq n}\left(R_{in}+R_{ni}\right)|\alpha_{i}|^{2}\alpha_{n} + 2\sum^{++-}_{ijk}{}^{'} S^{++-}_{ijkn}\alpha_{i}\alpha_{j}\bar{\alpha}_{k}
	\,,
\end{multline}
where we used \eqref{eq:22.03.15_08} and the symmetry of the $\alpha$ term to factor out $S^{++-}_{ijkn}$. Next, introducing
\begin{equation}
	\label{eq:22.03.16_06}
	R^{A}_{il} = \frac{1}{2}\left(R_{il} - R_{li}\right)\,,
\end{equation}
we rewrite \eqref{eq:22.03.16_04} as
\begin{equation}
	\label{eq:22.03.16_07}
	\frac{\partial V}{\partial \bar{\alpha}_{n}} = 2T_{n}|\alpha_{n}|^{2}\alpha_{n} + 2\sum_{i\neq n}R_{in}|\alpha_{i}|^{2}\alpha_{n} + 2\sum^{++-}_{ijk}{}^{'} S^{++-}_{ijkn}\alpha_{i}\alpha_{j}\bar{\alpha}_{k} - 2 \sum_{i}R^{A}_{in}|\alpha_{i}|^{2}\alpha_{n}\,,
\end{equation}
note $R^{A}_{ii}=0$. Therefore, using the equation of motion \eqref{eq:22.03.15_01}, we can write
\begin{equation}
	\label{eq:22.03.16_08}
	\frac{1}{2}\frac{\partial V}{\partial \bar{\alpha}_{l}} = 2i\omega_{l}\dot{\alpha}_{l} - \sum_{i}R^{A}_{il}|\alpha_{i}|^{2}\alpha_{l}
	\,.
\end{equation}
Next, we compute the time derivative of $V$. By using \eqref{eq:22.03.16_08} and its complex conjugate, we obtain
\begin{equation}
	\label{eq:22.03.16_09}
	\frac{\diff{} V}{\diff{\tau}} = \sum_{j} \left(\frac{\partial V}{\partial \alpha_{j}}\dot{\alpha}_{j} + \frac{\partial V}{\partial \bar{\alpha}_{j}}\dot{\bar{\alpha}}_{j}\right) = -2 \sum_{ij}R^{A}_{ij}|\alpha_{i}|^{2}\frac{\diff{}}{\diff{\tau}}|\alpha_{j}|^{2}\,.
\end{equation}
An explicit calculation using the formulas in Sec.~\ref{sec:InteractionCoefficients} gives
\begin{equation}
	\label{eq:22.03.16_10}
	R^{A}_{ij} = \omega_{j}^{2}\tilde{V}_{i} - \omega_{i}^{2}\tilde{V}_{j}\,, \quad \tilde{V}_{i}=4\omega_{i}^{2}\left(\Ci(2\omega_{i})-\log\omega_{i}\right)\,.
\end{equation}
This, in turn, allows us to rewrite \eqref{eq:22.03.16_09} as
\begin{multline}
	\label{eq:22.03.16_11}
	\frac{\diff{} V}{\diff{\tau}} = -2 \sum_{ij}\left(\omega_{j}^{2}\tilde{V}_{i} - \omega_{i}^{2}\tilde{V}_{j}\right)|\alpha_{i}|^{2}\frac{\diff{}}{\diff{\tau}}|\alpha_{j}|^{2} 
	\\
	= -2 \sum_{ij}\tilde{V}_{i}|\alpha_{i}|^{2}\frac{\diff{}}{\diff{\tau}}\left(\omega_{j}^{2}|\alpha_{j}|^{2}\right) + 2 \sum_{ij}\omega_{i}^{2} |\alpha_{i}|^{2}\frac{\diff{}}{\diff{\tau}}\left(\tilde{V}_{j}|\alpha_{j}|^{2}\right)
	\\
	= -2\sum_{i}\tilde{V}_{i}|\alpha_{i}|^{2}\frac{\diff{}}{\diff{\tau}}E + 2E\frac{\diff{}}{\diff{\tau}}\sum_{j}\tilde{V}_{j}|\alpha_{j}|^{2}
	\,.
\end{multline}
Since $E$ is conserved, we conclude
\begin{equation}
	\label{eq:22.03.16_12}
	\frac{\diff{}}{\diff{\tau}}\left(V - 2E\sum_{j}\tilde{V}_{j}|\alpha_{j}|^{2}\right) = 0\,.
\end{equation}
Thus, we find the third conserved quantity $H$ of the system \eqref{eq:22.03.15_01}:
\begin{equation}
	\label{eq:22.03.16_13}
	H = \frac{1}{2}V - E\sum_{j}\tilde{V}_{j}|\alpha_{j}|^{2}
	\,,
\end{equation}
analogous to the asymptotically AdS case \cite{Craps.2015}.

In summary, although the coefficients in the system \eqref{eq:22.03.15_01} do not share the same symmetries as the coefficients in the corresponding resonant system for scalar perturbations of AdS, we find that the $++-$ system has (at least) three conserved quantities. However, as the coefficients $S^{++-}_{ijkl}$ do not satisfy the required conditions, the additional quantity found to be conserved in the case of AdS \cite{Biasi.2019} is not preserved along the flow generated by \eqref{eq:22.03.15_01}.

\subsubsection{The full resonant system}

Due to the presence of the $+++$ and $+--$ terms, the sum $J$ is no longer preserved by the flow \eqref{eq:21.08.02_09}. However, the full resonant system has at least two conserved quantities: $E$ and a generalization of \eqref{eq:22.03.16_11}, as will be demonstrated below. The derivation builds on the analysis of the $++-$ system.

First, we show that $E=\sum_{i}\omega_{i}^{2}|\alpha_{i}|^{2}$ is conserved. Using \eqref{eq:21.08.02_09} and its complex conjugate, we rewrite the time derivative of $E$ as
\begin{multline}
	\label{eq:22.03.16_14}
	\frac{\diff{}}{\diff{\tau}}E = \sum_{l}\omega_{l}^{2}\left(\dot{\alpha}_{l}\bar{\alpha}_{l}+\alpha_{l}\dot{\bar{\alpha}}_{l}\right)
	= -\frac{i}{2}\left[
	\sum^{+++}_{ijkl}\omega_{l}S^{+++}_{ijkl}\left(\alpha_{i}\alpha_{j}\alpha_{k}\bar{\alpha}_{l} - \bar{\alpha}_{i}\bar{\alpha}_{j}\bar{\alpha}_{k}\alpha_{l}\right)
	\right.
	\\
	\left.
	+ \sum^{++-}_{ijkl}\omega_{l}S^{++-}_{ijkl}\left(\alpha_{i}\alpha_{j}\bar{\alpha}_{k}\bar{\alpha}_{l} - \bar{\alpha}_{i}\bar{\alpha}_{j}\alpha_{k}\alpha_{l}\right)
	+ \sum^{+--}_{ijkl}\omega_{l}S^{+--}_{ijkl}\left(\alpha_{i}\bar{\alpha}_{j}\bar{\alpha}_{k}\bar{\alpha}_{l} - \bar{\alpha}_{i}\alpha_{j}\alpha_{k}\alpha_{l}\right)
	\right]\,.
\end{multline}
The middle sum vanishes, as was demonstrated for the $++-$ system. We rewrite the $+--$ term so that it can be combined with the $+++$ term. We have
\begin{multline}
	\label{eq:22.03.16_15}
\sum^{+--}_{ijkl}\omega_{l}S^{+--}_{ijkl}\left(\alpha_{i}\bar{\alpha}_{j}\bar{\alpha}_{k}\bar{\alpha}_{l} - \bar{\alpha}_{i}\alpha_{j}\alpha_{k}\alpha_{l}\right) = 
\sum_{\substack{ijkl \\ l-j-k-2=i}}\omega_{i}S^{+--}_{ljki}\left(\bar{\alpha}_{i}\bar{\alpha}_{j}\bar{\alpha}_{k}\alpha_{l} - \alpha_{i}\alpha_{j}\alpha_{k}\bar{\alpha}_{l}\right) 
	\\
	= \sum^{+++}_{ijkl}\omega_{i}S^{+--}_{ljki}\left(\bar{\alpha}_{i}\bar{\alpha}_{j}\bar{\alpha}_{k}\alpha_{l} - \alpha_{i}\alpha_{j}\alpha_{k}\bar{\alpha}_{l}\right)\,,
\end{multline}
where by using the permutation $i\leftrightarrow l$, we obtained the $+++$ resonant condition within the sum. Under this manipulation, both the $+++$ and $+--$ sums in \eqref{eq:22.03.16_14} contain the same resonant condition, thus they can be put together. This way, we get
\begin{multline}
	\label{eq:22.03.16_16}
	2i\frac{\diff{}}{\diff{\tau}}E = \sum^{+++}_{ijkl}\omega_{l}S^{+++}_{ijkl}\left(\alpha_{i}\alpha_{j}\alpha_{k}\bar{\alpha}_{l} - \bar{\alpha}_{i}\bar{\alpha}_{j}\bar{\alpha}_{k}\alpha_{l}\right)
	+\sum^{+--}_{ijkl}\omega_{l}S^{+--}_{ijkl}\left(\alpha_{i}\bar{\alpha}_{j}\bar{\alpha}_{k}\bar{\alpha}_{l} - \bar{\alpha}_{i}\alpha_{j}\alpha_{k}\alpha_{l}\right)
	\\
	= \sum^{+++}_{ijkl}\omega_{l}S^{+++}_{ijkl}\left(\alpha_{i}\alpha_{j}\alpha_{k}\bar{\alpha}_{l} - \bar{\alpha}_{i}\bar{\alpha}_{j}\bar{\alpha}_{k}\alpha_{l}\right)
	+ \sum^{+++}_{ijkl}\omega_{i}S^{+--}_{ljki}\left(\bar{\alpha}_{i}\bar{\alpha}_{j}\bar{\alpha}_{k}\alpha_{l} - \alpha_{i}\alpha_{j}\alpha_{k}\bar{\alpha}_{l}\right)
	\\
	= \sum^{+++}_{ijkl}\left(\omega_{l}S^{+++}_{ijkl}-\omega_{i}S^{+--}_{ljki}\right)\left(\alpha_{i}\alpha_{j}\alpha_{k}\bar{\alpha}_{l} - \bar{\alpha}_{i}\bar{\alpha}_{j}\bar{\alpha}_{k}\alpha_{l}\right)\,.
\end{multline}
Using the fact that we can permute the summation indices $i,j,k$ in this sum, we can rewrite \eqref{eq:22.03.16_16} as
\begin{equation}
	\label{eq:22.03.16_17}
	2i\frac{\diff{}}{\diff{\tau}}E 
	= \sum^{+++}_{ijkl}W_{ijkl}\left(\alpha_{i}\alpha_{j}\alpha_{k}\bar{\alpha}_{l} - \bar{\alpha}_{i}\bar{\alpha}_{j}\bar{\alpha}_{k}\alpha_{l}\right)\,,
\end{equation}
where
\begin{multline}
	\label{eq:22.03.16_18}
	W_{ijkl} = \frac{\omega_{l}}{3}\left(S^{+++}_{ijkl} + S^{+++}_{jikl} + S^{+++}_{kjil}\right)
	- \frac{1}{3}\left(\omega_{i}S^{+--}_{ljki} + \omega_{j}S^{+--}_{likj} + \omega_{k}S^{+--}_{ljik}\right)
	\,.
\end{multline}
From the property of $W_{ijkl}$
\begin{equation}
	\label{eq:22.03.16_19}
	W_{ijkl} = - W_{ikjl}\,,	
\end{equation}
which holds for $i+j+k+2=l$, we get
\begin{equation}
	\label{eq:22.03.16_20}
	\frac{\diff{}}{\diff{\tau}}E = 0\,.
\end{equation}

Next, we look for a conserved quantity $H$ analogous to the $++-$ system. We define
\begin{equation}
	\label{eq:22.03.16_21}
	X = \sum^{+++}_{ijkl}S^{+++}_{ijkl}\alpha_{i}\alpha_{j}\alpha_{k}\bar{\alpha}_{l}\,,
\end{equation}
and
\begin{equation}
	\label{eq:22.03.16_22}
	Z = \sum^{+--}_{ijkl}S^{+--}_{ijkl}\alpha_{i}\bar{\alpha}_{j}\bar{\alpha}_{k}\bar{\alpha}_{l}
	\,.
\end{equation}
Then
\begin{equation}
	\label{eq:22.03.16_23}
	\frac{\partial X}{\partial \bar{\alpha}_{n}} = \sum^{+++}_{ijk}S^{+++}_{ijkn}\alpha_{i}\alpha_{j}\alpha_{k}\,,
\end{equation}
and
\begin{equation}
	\label{eq:22.03.16_24}
	\frac{\partial Z}{\partial \bar{\alpha}_{n}} = \sum^{+--}_{ijk}\left(S^{+--}_{ijkn}+S^{+--}_{ijnk}+S^{+--}_{inkj}\right)\alpha_{i}\bar{\alpha}_{j}\bar{\alpha}_{k}
	\,.
\end{equation}
Using the symmetry of the $\alpha$ factor in \eqref{eq:22.03.16_24}, we have
\begin{equation}
	\label{eq:22.03.16_25}
	\frac{\partial Z}{\partial \bar{\alpha}_{n}} = \sum^{+--}_{ijk}\left[S^{+--}_{ijkn} + \frac{1}{2}\left(S^{+--}_{ijnk}+S^{+--}_{iknj}+S^{+--}_{inkj}+S^{+--}_{injk}\right)\right]\alpha_{i}\bar{\alpha}_{j}\bar{\alpha}_{k}\,.
\end{equation}
Now we use the identity for $S^{+--}_{ijkl}$
\begin{equation}
	\label{eq:22.03.16_26}
	0 = S^{+--}_{ijkl} + S^{+--}_{ikjl} - \frac{1}{2}\left(S^{+--}_{ijlk} + S^{+--}_{iklj} + S^{+--}_{ilkj} + S^{+--}_{iljk}\right)\,,
\end{equation}
for $i-j-k-2=l$ to rewrite \eqref{eq:22.03.16_24} as
\begin{equation}
	\label{eq:22.03.16_27}
	\frac{\partial Z}{\partial \bar{\alpha}_{n}} = 3\sum^{+--}_{ijk}S^{+--}_{ijkn}\alpha_{i}\bar{\alpha}_{j}\bar{\alpha}_{k}\,.
\end{equation}

Using $X$, $Z$, and the result for $V$ derived in \eqref{eq:22.03.16_07}, we can rewrite the resonant system \eqref{eq:21.08.02_09} as
\begin{equation}
	\label{eq:22.03.16_28}
	2i\omega_{l}\dot{\alpha}_{l} = \frac{\partial W}{\partial \bar{\alpha}_{l}} + \sum_{i}R^{A}_{il}|\alpha_{i}|\alpha_{l}\,,
\end{equation}
where we defined
\begin{equation}
	\label{eq:22.03.16_29}
	W = X + \frac{1}{2}V + \frac{1}{3}Z\,.
\end{equation}
Following the same steps as in the derivation of $H$ for the $++-$ system, we show
\begin{equation}
	\label{eq:22.03.16_30}
	\frac{\diff{}}{\diff{\tau}}W = - \sum_{ij}R^{A}_{ij}|\alpha_{i}|^{2}\frac{\diff{}}{\diff{\tau}}|\alpha_{j}|^{2}
	\,,
\end{equation}
cf. \eqref{eq:22.03.16_09}. From the explicit form of $R^{A}_{il}$ given in \eqref{eq:22.03.16_10} and the constancy of $E$, it then follows that
\begin{equation}
	\label{eq:22.03.16_31}
	H = W - E\sum_{j}\tilde{V}_{j}|\alpha_{j}|^{2}
	\,,
\end{equation}
is conserved. We note that by dropping the $S^{+++}$ and $S^{+--}$, which amounts to setting $X=0=Z$ in \eqref{eq:22.03.16_31}, we recover the formula \eqref{eq:22.03.16_13} for the $++-$ system.

Remark: below we propose an alternative expression for computing \eqref{eq:22.03.16_31}, i.e.,
\begin{equation}
	\label{eq:22.07.05_01}
	H = \frac{1}{2}\re\left(X+V+Z\right) - E\sum_{j}\tilde{V}_{j}|\alpha_{j}|^{2}
	\,.
\end{equation}
To show that these definitions agree, we need the result
\begin{equation}
	\label{eq:22.07.05_02}
	X-\frac{1}{3}\bar{Z} = 0
	\,,
\end{equation}
which we prove below. Using the definitions \eqref{eq:22.03.16_21}-\eqref{eq:22.03.16_22}, we have
\begin{multline}
	\label{eq:22.07.05_03}
	X-\frac{1}{3}\bar{Z} = \sum^{+++}_{ijkl}S^{+++}_{ijkl}\alpha_{i}\alpha_{j}\alpha_{k}\bar{\alpha}_{l} 
	- \frac{1}{3}\sum^{+--}_{ijkl}S^{+--}_{ijkl}\bar{\alpha}_{i}\alpha_{j}\alpha_{k}\alpha_{l}
	\\
	= \sum_{ijkl}^{+++}\left(S^{+++}_{ijkl}-\frac{1}{3}S^{+--}_{ijkl}\right)\alpha_{i}\alpha_{j}\alpha_{k}\bar{\alpha}_{l}
	\,,
\end{multline}
where we used the renaming of indices $i\leftrightarrow l$ to factor out the product of $\alpha$'s. Note that the $+--$ becomes $+++$ then. Defining
\begin{equation}
	\label{eq:22.07.05_04}
	U_{ijkl} = S^{+++}_{ijkl}-\frac{1}{3}S^{+--}_{ijkl}
	\,,
\end{equation}
we rewrite \eqref{eq:22.07.05_03} as 
\begin{equation}
	\label{eq:22.07.05_05}
	X-\frac{1}{3}\bar{Z} = \sum_{ijkl}^{+++}U_{(ijk)l}\alpha_{i}\alpha_{j}\alpha_{k}\bar{\alpha}_{l}
	\,.
\end{equation}
From $U_{(ijk)l}=0$, the vanishing of \eqref{eq:22.07.05_02} follows immediately. Moreover, from \eqref{eq:22.07.05_02}, we get
\begin{equation}
	\label{eq:22.07.05_06}
	X + \frac{1}{3}Z - \left(\bar{X} + \frac{1}{3}\bar{Z}\right) = X-\frac{1}{3}\bar{Z} - \left(\bar{X}-\frac{1}{3}Z\right) = 0
	\,,
\end{equation}
so
\begin{equation}
	\label{eq:22.07.05_07}
	X + \frac{1}{3}Z \in \mathbb{R}
	\,.
\end{equation}
We also have
\begin{equation}
	\label{eq:22.07.05_08}
	\re\left(X-\frac{1}{3}Z\right) = \frac{1}{2}\left(X-\frac{1}{3}Z + \bar{X}-\frac{1}{3}\bar{Z}\right)
	= \frac{1}{2}\left(X-\frac{1}{3}\bar{Z} + \overline{\left(X-\frac{1}{3}\bar{Z}\right)} \right) = 0
	\,.
\end{equation}
Now, subtracting \eqref{eq:22.03.16_31} from \eqref{eq:22.07.05_01}, we obtain
\begin{multline}
	\label{eq:22.07.05_09}
	\left(\frac{1}{2}\re\left(X+V+Z\right) - E\sum_{j}\tilde{V}_{j}|\alpha_{j}|^{2}\right) - \left(X + \frac{1}{2}V + \frac{1}{3}Z - E\sum_{j}\tilde{V}_{j}|\alpha_{j}|^{2} \right)
	\\
	=  \frac{1}{2}\re(X+Z) - \left(X+\frac{1}{3}Z\right)
	\,.
\end{multline}
Using \eqref{eq:22.07.05_07} and \eqref{eq:22.07.05_08}, one shows that the difference in \eqref{eq:22.07.05_09} vanishes. Consequently, the definitions \eqref{eq:22.03.16_31} and \eqref{eq:22.07.05_01} agree. Using \eqref{eq:22.07.05_01} instead of \eqref{eq:22.03.16_31} in numerical calculations has a great advantage, since the first term in the expression can be efficiently computed as a dot product of $\alpha_{l}$ and $\diff\alpha_{l}/\diff{\tau}$. Additionally, this way of computing $H$ reduces the rounding errors, which usually get large at late times of the evolution of singular solutions.

The discussion of conserved quantities was independent of the residual gauge choice, which does not affect the symmetries of the interaction coefficients used in the analysis. It follows that the same functional form of the conserved quantities holds both in the origin gauge $\delta(t,r=0)=0$ and the boundary gauge $\delta(t,r=1)=0$.
The boundary time gauge is discussed in Appendix~\ref{sec:BoundaryTimeGauge}.

We note that in the boundary time gauge, where the resonant system is Hamiltonian, the conserved quantities $E$, $J$ (for the $++-$ system only) and $H$ follow from the respective symmetries ($\theta,\tau_{0}\in\mathbb{R}$)
\begin{equation}
	\label{eq:22.08.31_01}
	\alpha_{l}(\tau) \rightarrow e^{il\theta}\alpha_{l}(\tau)\,,
	\quad
	\alpha_{l}(\tau) \rightarrow e^{i\theta}\alpha_{l}(\tau)\,,
	\quad
	\alpha_{l}(\tau) \rightarrow \alpha_{l}(\tau-\tau_{0})
	\,.
\end{equation}
The extra resonant terms present in the full resonant system are responsible for the lack of the global phase shift symmetry $\alpha_{l}(\tau) \rightarrow e^{i\theta}\alpha_{l}(\tau)$, from which it follows that $J$ is not conserved by the flow \eqref{eq:21.08.02_09}.

\section{Results}
\label{sec:Results}

\subsection{Numerical solution}

We truncate the system \eqref{eq:21.08.02_09}, i.e. we solve it for the first $N$ modes, thus the state vector is $\{\alpha_{i}\}_{i=0,\ldots,N-1}$, and the sums on the rhs are truncated accordingly. 
We point out that due to the structure of the equations in \eqref{eq:21.08.02_09}, the numerical solution of the resonant system poses several difficulties. Although the interaction coefficients are explicit in the case at hand (this fact makes the system particularly interesting from the theoretical point of view) and their computation is not particularly involved,\footnote{To evaluate the functions \eqref{eq:22.07.23_03} we use a variant of the algorithm proposed in \cite{Rowe.2015}.} the time integration of large systems is a limiting factor, particularly in the boundary time gauge, see Appendix \ref{sec:BoundaryTimeGauge}.
The truncation should be sufficiently large so that the errors do not spoil the numerical solution. 
In practice, however, $N$ cannot be too big, as then computations quickly become very expensive. The memory required to store the interaction coefficients is $\mathcal{O}(N^{3})$, while the number of floating point operations is $\mathcal{O}(N^{4})$ (computing the rhs requires $\mathcal{O}(N^{3})$ operations whereas the integration time step should be $\mathcal{O}(N^{-1})$ to accurately resolve rapid oscillations of highest modes, cf. Eq.~\eqref{eq:22.03.21_08} below).

In choosing $N$, we make a compromise between the computational cost and the truncation errors. We carefully tested our results so that the artefacts were minimized. We drop roughly half of the highest modes from the analysis, and we make sure that we do not consider data corresponding to the integration past the singularity formation. As it turns out, the systems we studied were still too small to reach conclusive answers (see the discussion below). Therefore the analysis presented below, which aims at understanding the singular solution, is less rigorous than the content of the preceding sections.

We consider the initial conditions corresponding to the initial values used for the Einstein equations\footnote{Modulo the rescaling \eqref{eq:22.07.20_01}, which is introduced so that the integration interval of the resonant system equations is not too large; the scaling is then adjusted when producing plots and comparing with the solution to the Einstein-Scalar field system.} in Eq.~\eqref{eq:20.12_01}.
We note in passing that the data consisting of a finite number of modes (e.g. the two-mode initial data considered in \cite{Bizoń.2015}) would lead to a nonsmooth solution for the Einstein-Scalar field system due to the violation of the corner conditions mentioned in Sec.~\ref{sec:LinearProblem}. Therefore, we do not discuss such configurations here. Nevertheless, we observe a similar singular evolution for initial configurations consisting of the two lowest modes with equal energy \eqref{eq:22.06.21_01}, and no singular solution when a significant portion of the energy is concentrated in one of the initially excited modes, in accordance with similar experiments in the AdS case. Due to the extra resonant interactions, for one-mode initial data, other modes get excited in the evolution \cite{Maliborski.2015}. This is contrary to the $++-$ system, where a single mode is a solution to the resonant system. However, such data also does not lead to a singular evolution. Instead, it may be considered as a perturbation of a time-periodic solution bifurcating from the corresponding eigenfrequency \cite{Maliborski.2015}.

\begin{figure}[!t]
	\includegraphics[width=0.495\textwidth]{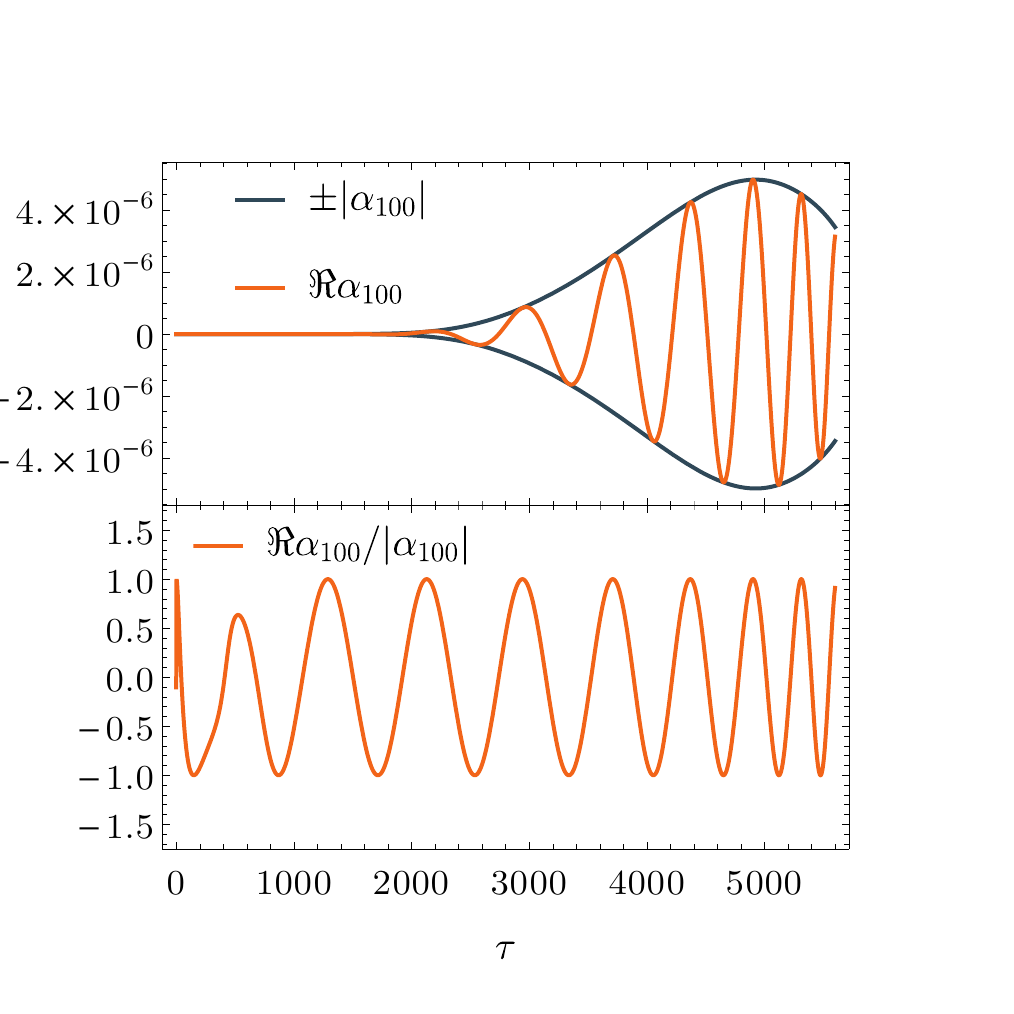}
	\includegraphics[width=0.495\textwidth]{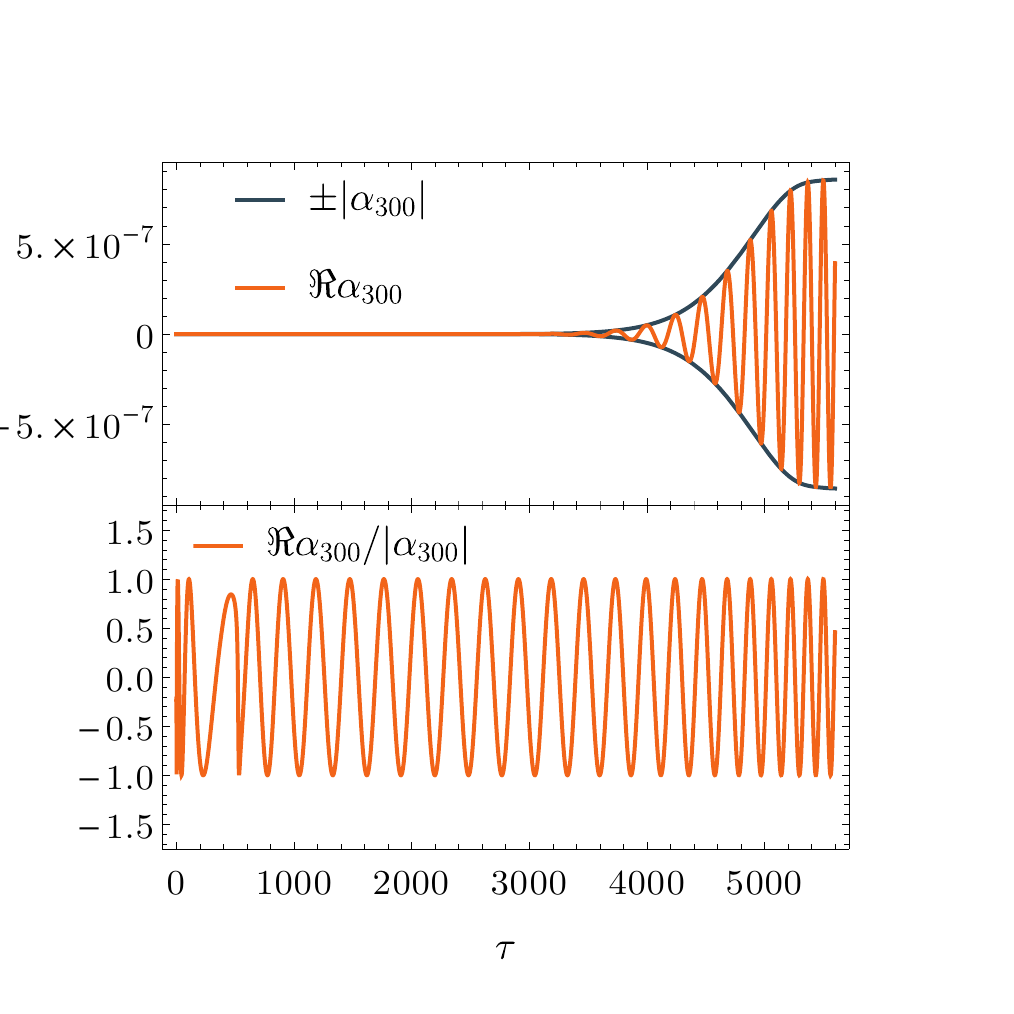}
	\caption{Time evolution of sample modes. We observe qualitatively similar behaviour for other modes. The frequency increases linearly with the mode index.}
	\label{fig:RS_AmplitudeAndRe_Set1}
\end{figure}

\begin{figure}[!t]
	\includegraphics[width=0.47\textwidth]{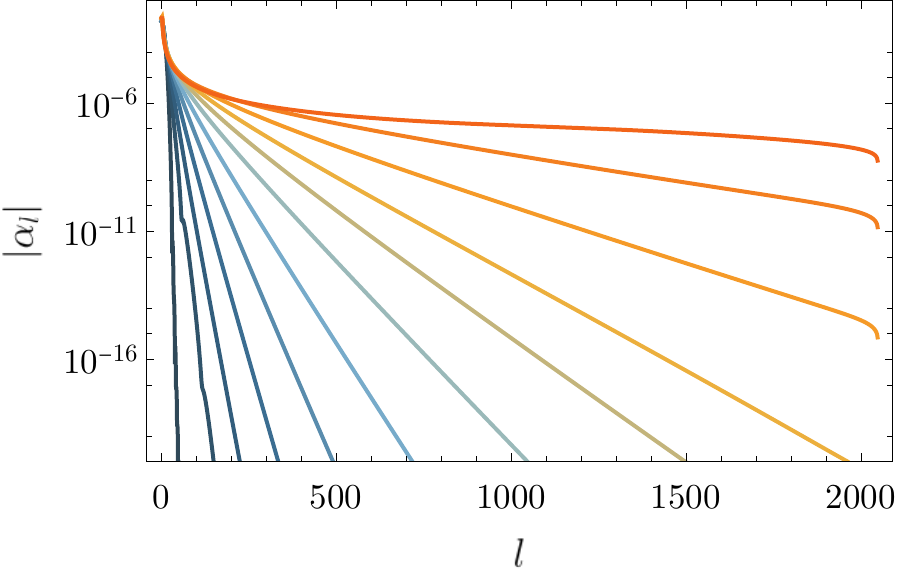}
	\hspace{5ex}
	\includegraphics[width=0.43\textwidth]{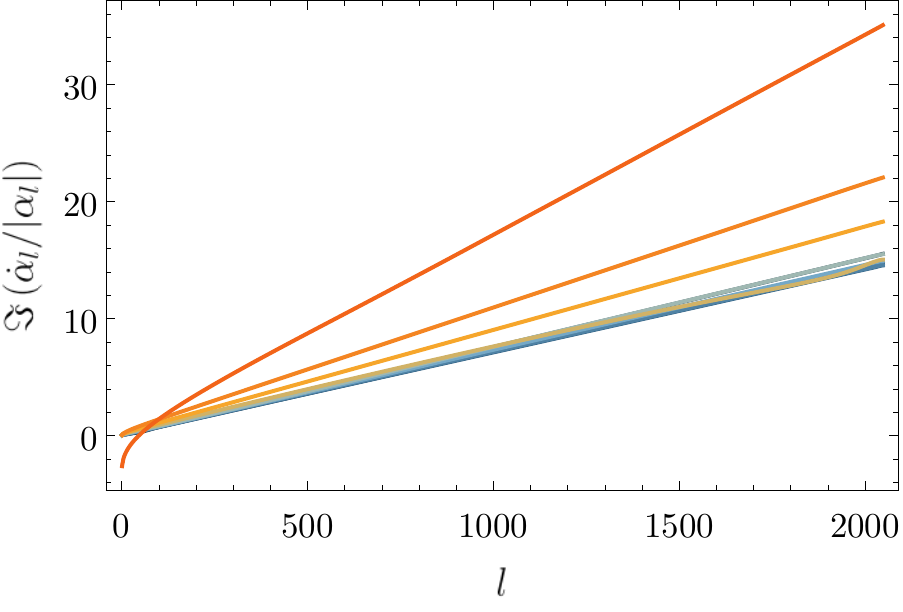}
	\caption{Typical behavior of mode amplitudes $A_{l}=|\alpha_{l}|$ and the phase derivatives $\dot{B}_{l}=\Im(\dot{\alpha}_{l}/|\alpha_{l}|)$ for initial data experiencing singular behavior. The slope of the amplitude spectra decreases monotonically with $\tau$. At later times, before the effects of truncation of the resonant system become visible, the spectrum unfolds a polynomial tail $A_{l}\sim l^{-\beta}$. The phases get synchronized, i.e. $B_{l}\sim l$, almost immediately and stay so for all times. Time is color coded and increases from bluish to reddish colors.}
	\label{fig:RSAmplitudePhaseEvolution}
\end{figure}

It is convenient to analyze the solutions in terms of the mode amplitudes $A_{l}$ and phases $B_{l}$ defined by
\begin{equation}
	\label{eq:22.06.28_01}
	\alpha_{l}(\tau) = A_{l}(\tau) e^{i B_{l}\left(\tau\right)}
	\,.
\end{equation}
As in \cite{Bizoń.2015}, we see a steady, monotonic growth of amplitudes and almost immediate synchronization of phases, i.e., $B_{l}(\tau) \sim l$.
For later times, we observe that the synchronization of phases persists, and at the same time, the frequency of oscillations increases during the evolution. Although not so strong as for the AdS$_{5}$ case \cite{Bizoń.2015,Deppe.20191vp}, the growth of $B_{l}(\tau)$ with increasing $\tau$ is noticeable. At the same time the amplitudes of higher modes get substantially excited, see Fig.~\ref{fig:RS_AmplitudeAndRe_Set1}, and at later times a polynomial tail unfolds, i.e., for $l\gg 1$ we have $A_{l}\sim l^{-\beta}$ with a universal exponent $\beta>0$, see Fig.~\ref{fig:RSAmplitudePhaseEvolution}. This asymptotic solution is analyzed in detail in the next section.

\begin{figure}
	\includegraphics[width=0.9\textwidth]{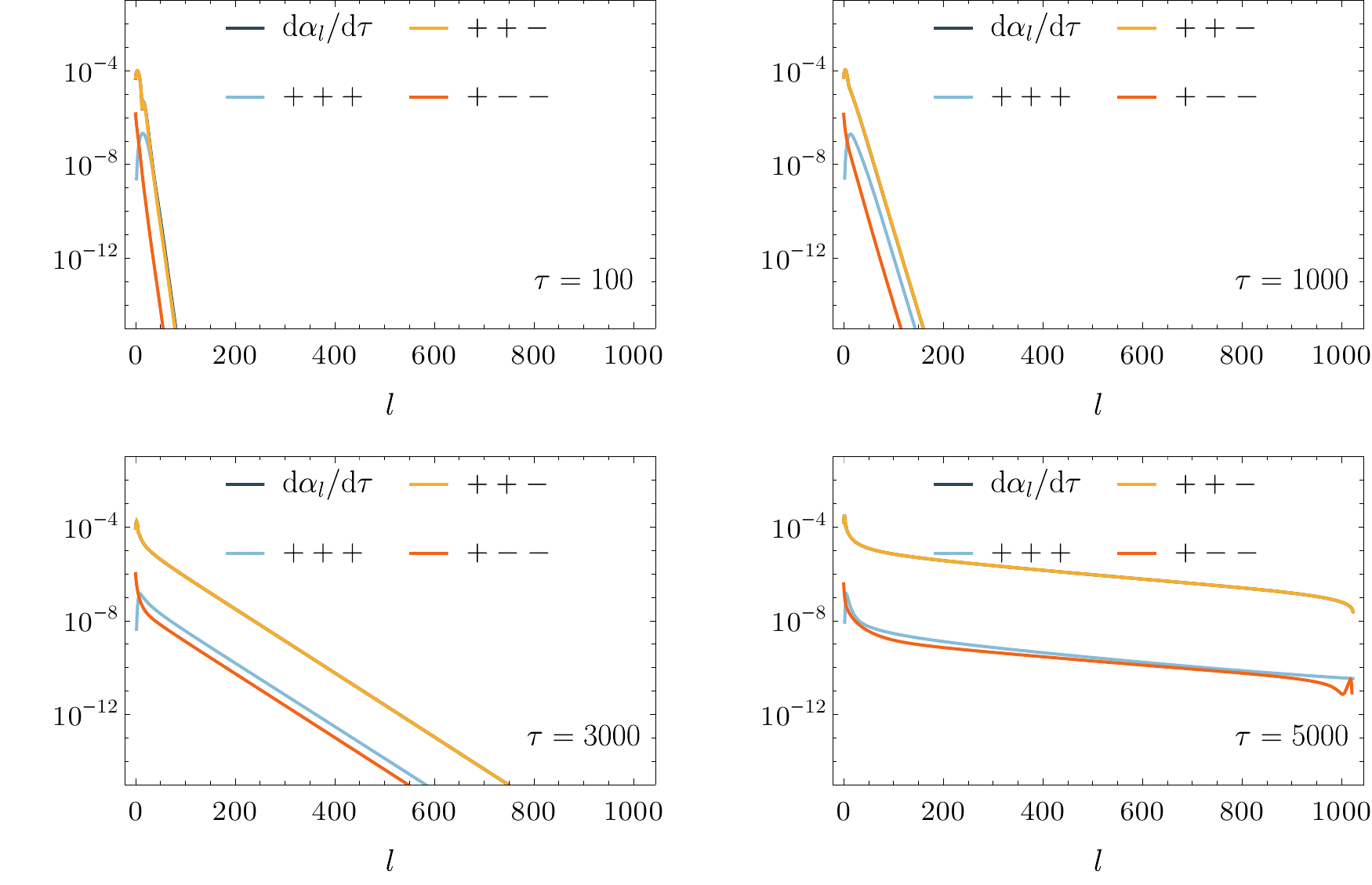}
	\caption{Snapshots from the evolution of the full resonant system. We plot the absolute value of time derivatives of $\alpha_{l}$, i.e. the rhs of Eq.~\eqref{eq:21.08.02_09} and compare it to the magnitude of each sum corresponding to a different resonant combination. Clearly, the rhs is dominated by the $++-$ term as the blue ($\diff{\alpha_{l}}/\diff{\tau}$) and yellow ($++-$) lines are indistinguishable on the scale of the plot. This is especially evident close to the time when the polynomial spectrum develops, where the contribution from the other resonant interactions $+++$ and $+--$ is two orders of magnitude smaller.}
\label{fig:RS_PPPvsPPMvsPMM_Set1}
\end{figure}

Interestingly, it turns out that in the evolution of initial data leading to a singular solution, the $+++$ and $+--$ resonances are subdominant, whereas the $++-$ resonances largely determine the evolution.\footnote{All resonant terms are essential for initial data that does not evolve toward a singular solution, e.g. single-mode initial data.} We illustrate this in Fig.~\ref{fig:RS_PPPvsPPMvsPMM_Set1} where we compare the magnitudes of the sums $+++$, $++-$ and $+--$ appearing on the rhs of \eqref{eq:21.08.02_09} during the evolution. The dominant role of the $++-$ resonance is evident. Almost from the very beginning, the evolution is driven by this resonant term. On the scale of the plot, the two lines corresponding to the $++-$ term and the rhs of \eqref{eq:21.08.02_09} are indistinguishable. This strongly suggests that both the full resonant system \eqref{eq:21.08.02_09} and the $++-$ system \eqref{eq:22.03.15_01} have singular solutions and that the latter serves as a good approximation to the former.

\begin{figure}
\includegraphics[width=0.9\textwidth]{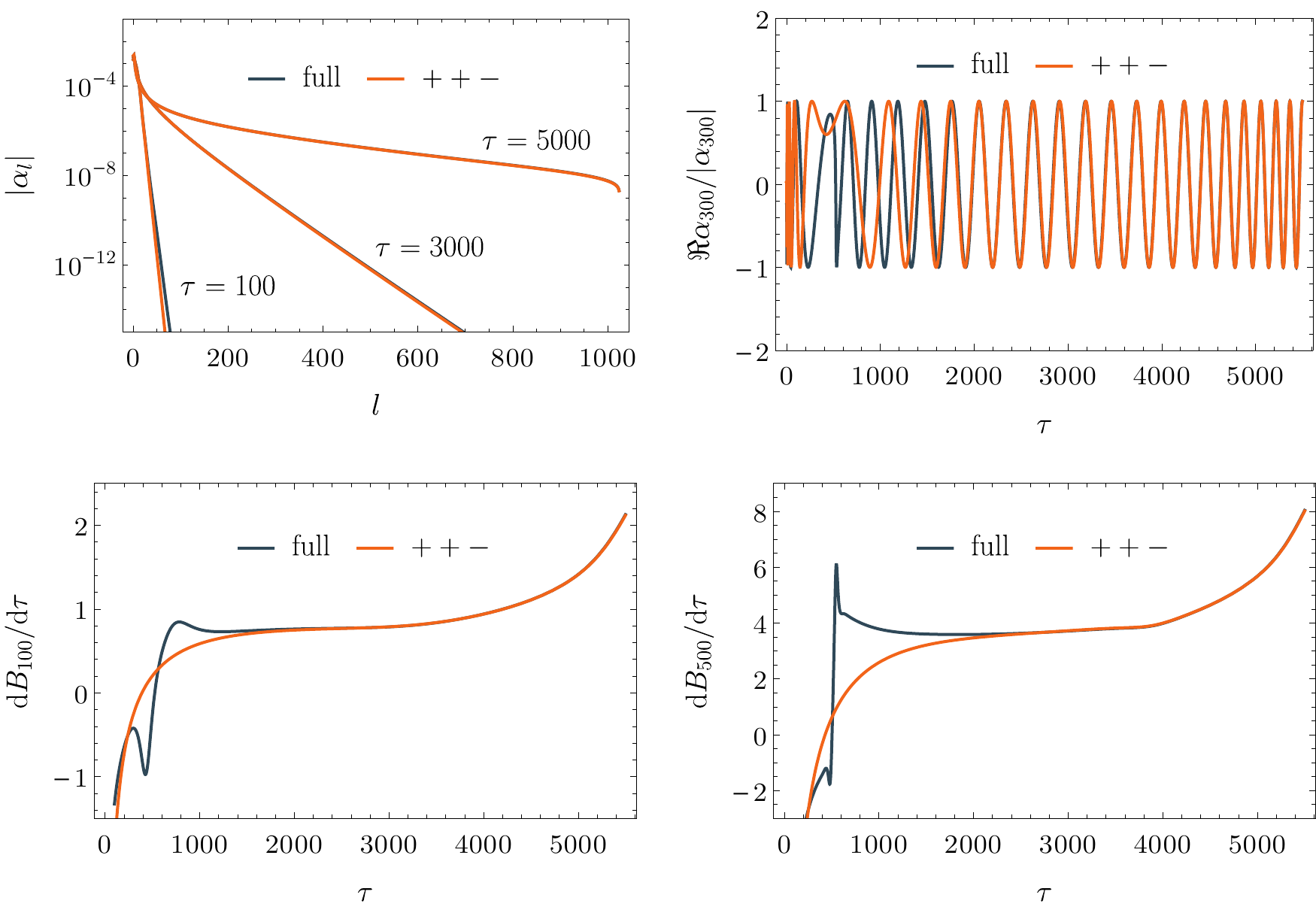}
\caption{Comparison of the solutions of the full resonant system \eqref{eq:21.08.02_09} and the $++-$ system \eqref{eq:22.03.15_01}. Both solutions start with the same initial data.
The amplitudes $|\alpha_{l}|$ of the two solutions agree so well that the blue (full) and orange ($++-$) lines coincide on the scale of the plot, especially at late times. Although significant differences in the initial stage of the evolution are visible on the plots of $\cos{B_{l}}$ and time derivatives of phases $\diff{B_{l}}/\diff{\tau}$, the solutions eventually get close to each other.}
\label{fig:RS_PPPvsPPMvsPMM2_Set1}
\end{figure}

To further test this hypothesis, we independently evolved the same initial data using the $++-$ system \eqref{eq:22.03.15_01}. The detailed comparison with the full system is presented in Fig.~\ref{fig:RS_PPPvsPPMvsPMM2_Set1}. Although the significance of the extra resonances is clearly observable at the early stages of the evolution, this is not the case at later times when the solutions are close to one another in the configuration space $\{\alpha_{i}\}_{i=0,\ldots,N-1}$. Therefore in the following analysis of the asymptotic solution, we focus on the $++-$ system \eqref{eq:22.03.15_01}.

\subsection{Asymptotic solution}
\label{sec:AsymptoticSolution}
First, we rewrite the system \eqref{eq:22.03.15_01} using amplitudes $A_{l}$ and phases $B_{l}$ introduced in \eqref{eq:22.06.28_01}. Then, the complex system \eqref{eq:22.03.15_02} is equivalent to the set of real equations
\begin{align}
	\label{eq:22.06.28_02}
	2\omega_{l}\frac{\diff}{\diff{\tau}}A_{l} &= \sum_{ijk}{}^{'}S^{++-}_{ijkl}A_{i}A_{j}A_{k}\sin\left(B_{i}+B_{j}-B_{k}-B_{l}\right)
	\,,
	\\
	\label{eq:22.06.28_03}
	-2\omega_{l}\frac{\diff}{\diff{\tau}}B_{l} &= T_{l}A_{l}^{2} + \sum_{i\neq l}R_{il}A_{i}^{2} + \sum_{ijk}{}^{'}S^{++-}_{ijkl}A_{i}A_{j}A_{k}\cos\left(B_{i}+B_{j}-B_{k}-B_{l}\right)
	\,.
\end{align}
To analyze the asymptotic solution, we use the analyticity strip method \cite{Bizoń.2013, Sulem.1983}. This amounts to writing the asymptotic ansatz for the amplitudes
\begin{equation}
	\label{eq:22.07.23_01}
	A_{l}(\tau) \sim l^{-\beta(\tau)} e^{-\rho(\tau)l}, \quad l \gg 1
	\,.
\end{equation}
Fits to the numerical data predict that the analyticity radius $\rho(\tau)$ tends to zero in some finite time $\tau_{*}$, which indicates that the solution of the resonant system \eqref{eq:22.03.15_01} becomes singular at $\tau_{*}$. 
Increasing the truncation parameter $N$, we observe a tendency for both $\beta$ and $\tau_{*}$ to grow and to approach limiting values. 
The run with the largest truncation of $N=2048$ modes gives $\beta\in (1.55,1.58)$ and $\tau_{*}\in (5491,5523)$, values close to the numbers read off from the solution of the Einstein-Scalar field system, cf. \eqref{eq:22.07.12_04} and \eqref{eq:22.07.12_02}.

The difficulty in determining the value of $\beta$ and the precise location of the singularity $\tau_{*}$ from the amplitude spectrum is caused by a combination of the fitting errors (fits are sensitive to the fitting interval) and the truncation error. In this model, we observe a slow convergence of the truncated resonant system to its infinite version, see \cite{Deppe.20191vp} for a similar observation in the AdS$_{4}$ case.
To minimize the systematic error, we only consider fitting intervals where the variation of the result is minimal and the truncation effects are the smallest (e.g. the right endpoint of the fitting interval is not higher than $N/2$ for an $N$ mode truncation). 

Following \cite{Bizoń.2013} (see also \cite{Deppe.20191vp}) we study the time evolution of frequencies of the singular solution, assuming for $\tau\lesssim \tau_{*}$ the asymptotic form \eqref{eq:22.07.23_01} with $\rho(\tau)=\rho_0(\tau_{*}-\tau)$, $\rho_{0}>0$. To find the asymptotic behavior of $\diff{B}_l/\diff{\tau}$ as $\tau\rightarrow\tau_{*}$, we drop the first term and the last sum in \eqref{eq:22.06.28_03} (i.e., the subdominant terms), and we consider:
\begin{equation}
	\label{eq:22.07.21_05}
	-2\omega_{l}\frac{\diff{B_{l}}}{\diff{\tau}} \approx \sum_{i\neq l} R_{il}A_{i}^2
	\,.
\end{equation}

Using the large argument expansion of the trigonometric integrals \eqref{eq:22.07.23_03}, we find
\begin{equation}
	\label{eq:22.07.25_01}
	\Si(2\omega_{l}) = \frac{\pi}{2}-\frac{1}{2\pi l} + \frac{1}{2\pi l^2} + \mathcal{O}\left(l^{-3}\right)
	\,,
	\quad
	\Ci(2\omega_{l}) = -\frac{1}{4\pi^{2}l^{2}} + \mathcal{O}\left(l^{-3}\right)
	\,,
\end{equation}
for large mode numbers $l\in\mathbb{N}$. From this, we get the asymptotic form of \eqref{eq:22.06.06_02}
\begin{equation}
	\label{eq:22.09.05_01}
	R_{il} \approx l^2\left(r_{5/2}i^{2}\log{i} + r_{2}i^{2} + r_{3/2}i\log{i} + r_{1}i \right)\,,
\end{equation}
where $r_{k}(l) = x_{k} + y_{k}/l + \mathcal{O}(l^{-2})$, and $x_{k},y_{k}$ are constants. Then the sum in \eqref{eq:22.07.21_05} can be computed using the ansatz \eqref{eq:22.07.23_01}. For large $l$, we get
\begin{multline}
	\label{eq:22.09.05_02}
	\frac{\diff{B_{l}}}{\diff{\tau}} \approx l \left[
	  c_{1} \partial_{\beta}\text{Li}_{2\beta-2}\left(e^{-2\rho}\right)
	+ c_{2} \text{Li}_{2\beta-2}\left(e^{-2\rho}\right)
	\right.
	\\
	\left.
	+ c_{3} \partial_{\beta}\text{Li}_{2\beta-1}\left(e^{-2\rho}\right)
	+ c_{4} \text{Li}_{2\beta-1}\left(e^{-2\rho}\right)
	\right]
	\,,
\end{multline}
where $\text{Li}_{s}(z)$ is the polylogarithm function \cite[\href{https://dlmf.nist.gov/25.12}{Sec.~25.12}]{NIST:DLMF}, and the $c_{i}$'s are functions of $l$ only, as follows from \eqref{eq:22.09.05_01}. From the asymptotic behaviour $B_{l}\sim l$ we get that $B_{i}+B_{j}-B_{k}-B_{l}\approx 0$ for the $++-$ resonant condition and also the consistency of the ansatz \eqref{eq:22.07.23_01}, since then both sides of Eq.~\eqref{eq:22.06.28_02} scale as $l^{2-\beta}$ for $l\rightarrow\infty$.

The leading order behavior of the polylogarithm functions appearing in \eqref{eq:22.09.05_02} for $\rho\rightarrow 0$ is
\begin{equation}
	\label{eq:22.09.05_03}
	\partial_{\beta}\text{Li}_{2\beta-2}\left(e^{-2\rho}\right) \sim \begin{cases}
		\rho^{2\beta-3}\log\rho\,, & \beta<3/2 \\
		\log^{2}\rho\,, & \beta=3/2 \\
		1\,, & \beta>3/2 \\
	\end{cases}
	\,,
	\quad\ 
	\text{Li}_{2\beta-2}\left(e^{-2\rho}\right) \sim \begin{cases}
		\rho^{2\beta-3}\,, & \beta<3/2 \\
		\log\rho\,, & \beta=3/2 \\
		1\,, & \beta>3/2 \\
	\end{cases}
\end{equation}
while both $\partial_{\beta}\text{Li}_{2\beta-1}\left(e^{-2\rho}\right)$ and $\text{Li}_{2\beta-1}\left(e^{-2\rho}\right)$ stay finite at $\rho=0$. Together, \eqref{eq:22.09.05_02} and \eqref{eq:22.09.05_03} imply the blowup of $\diff{B_{l}}/\diff{\tau}$ for $\beta\leq 3/2$ and the divergence of higher order derivatives as $\rho\rightarrow 0$, both for $\beta\leq 3/2$ and $\beta>3/2$.\footnote{For precise asymptotics one would need to consider a higher order expansion in \eqref{eq:22.09.05_03}, which we skip for clarity of presentation; however, see \eqref{eq:22.03.21_08}.}

As fitting the formula \eqref{eq:22.09.05_02} to the numerical data is particularly difficult (as depending on the fitting interval and the starting values, the fitting procedure does not converge, or the fits would not be unique), determining the precise value of $\beta$ from $\diff{B_{l}}/\diff{\tau}$ was not reliable (although we consistently obtained a value of $\beta>3/2$ for large $l$). Therefore we followed a different strategy.

\begin{figure}[!t]
	\includegraphics[width=0.5\textwidth]{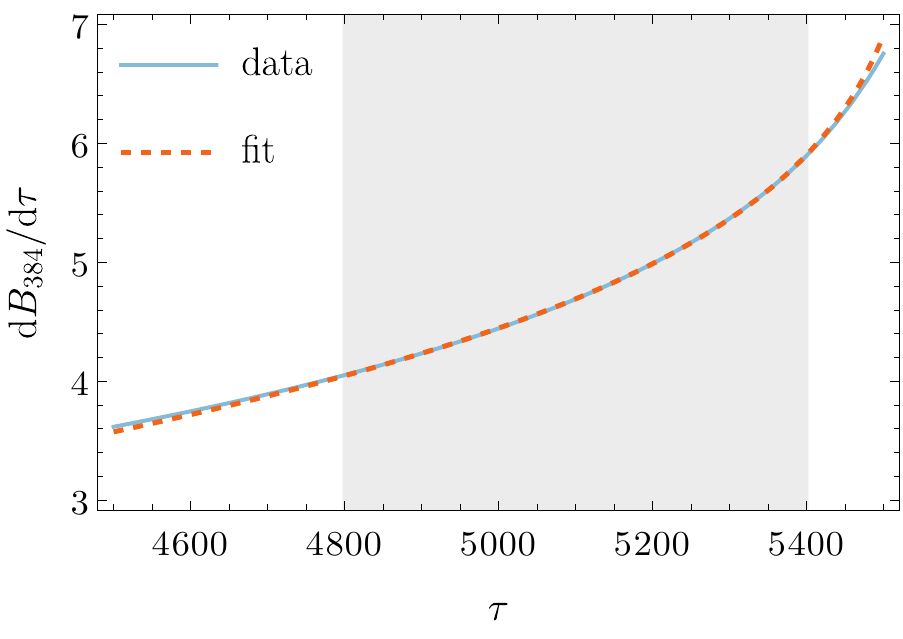}
	\caption{The asymptotic fit \eqref{eq:22.03.21_08} for the sample mode $l=384$. The shaded region indicates the fitting range. The solid line is the numerical data, while the dashed line is the fitted function.}
	\label{fig:fitAvsB_1}
\end{figure}

\begin{figure}[!t]
	\includegraphics[width=0.99\textwidth]{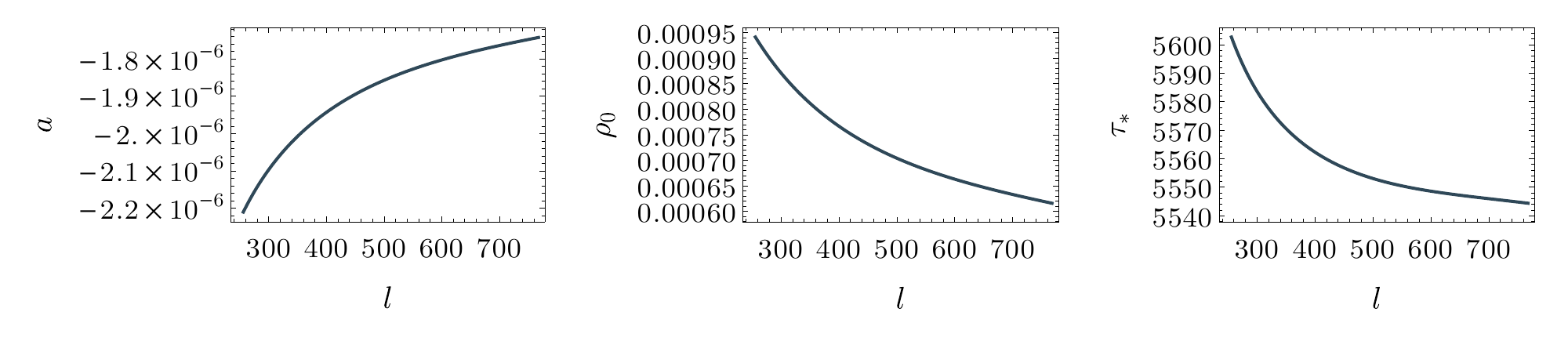}
	\caption{The dependence of the fitting parameters in \eqref{eq:22.03.21_08} on the mode number $l$. In the asymptotic regime when $\tau\approx\tau_{*}$ and $l\rightarrow\infty$, the fitting parameters should be independent of $l$.}
	\label{fig:fitAvsB_2}
\end{figure}

We fix the exponent to a value consistent with the energy spectrum $E_{j}(t\approx t_{AH})\sim j^{-6/5}$ observed prior to the AH formation in the Einstein-Scalar field system\footnote{Assuming $|\alpha_{l}|\sim l^{-\beta}$, it follows that at the leading order of $\varepsilon$, for $l\rightarrow\infty$, the energy spectrum is $E_{l}\sim l^{2(1-\beta)}$.}, i.e. we set $\beta = 8/5$ and fit the remaining parameters.
In this case, we obtain the following leading order behaviour for $\rho\rightarrow 0$ from \eqref{eq:22.09.05_02}:
\begin{equation}
    \label{eq:22.03.21_08}
	\frac{\diff{B_{l}}}{\diff{\tau}} \approx a\,l\left(\rho + b_{1} \rho^{1/5} + b_{2}\rho^{1/5}\log{\rho}\right)
	\,,
	\quad \Rightarrow \quad
	\frac{\diff{}^{2}B_{l}}{\diff{\tau}^{2}} \sim \rho^{-4/5}\log\rho
	\,,
\end{equation}
where $a$ and the $b_{i}$'s are constant. This predicts, blowup of higher order derivatives of $B_{l}$ at $\tau_{*}$, while the phases stay finite at the singularity. To test this prediction, we fitted \eqref{eq:22.03.21_08} to the numerical data for several different modes and obtained good agreement, see Fig.~\ref{fig:fitAvsB_1} for a representative result. However, fixing $\beta$ to other values close to $8/5$, e.g. within the range $(1.55,1.58)$ suggested by the amplitude spectra analysis, we get fits which do not vary considerably. Thus, instead of focusing on a fixed mode, we look at the dependence of the fitting parameters in (4) on the mode number.
It turns out that the variation of the overall amplitude $a$, the coefficient $\rho_{0}$, and the blowup time $\tau_{*}$ with respect to $l$ is relatively small (smaller than for other values of $\beta$ considered), cf. Fig.~\ref{fig:fitAvsB_2}, which further validates the approximation \eqref{eq:22.03.21_08}. Moreover, the value of $\beta=8/5$ is also favored by the analogous analysis of the system in boundary time gauge, discussion of which is delegated to the Appendix.

\begin{figure}[!t]
	\includegraphics[width=0.45\textwidth]{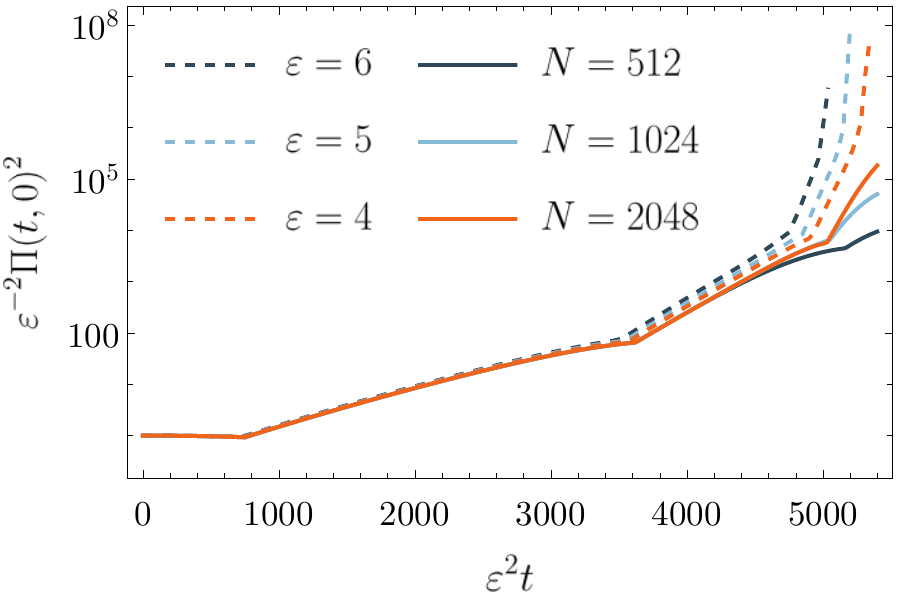}
	\hspace{1ex}
	\includegraphics[width=0.45\textwidth]{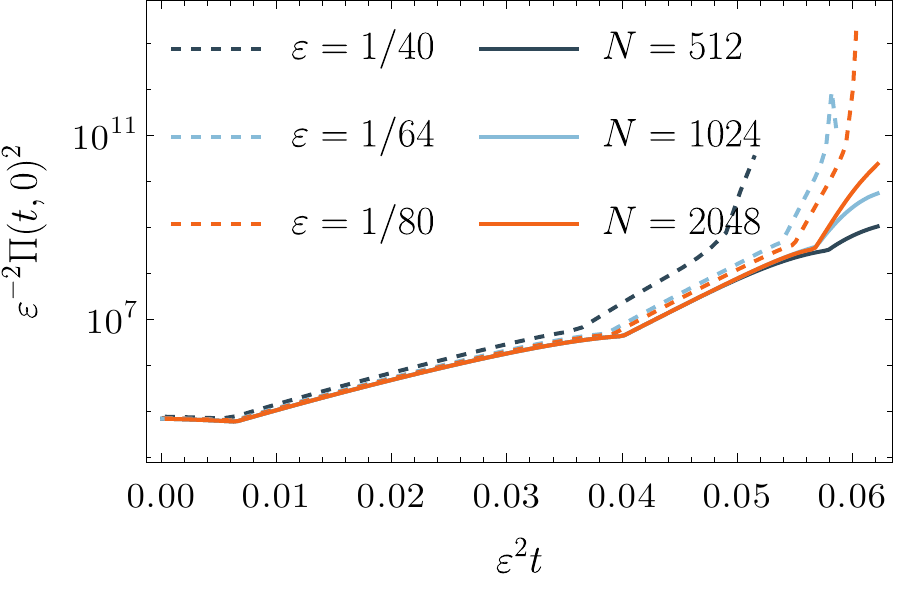}
	\caption{The rescaled Ricci scalar at the origin computed from the solution of the Einstein-Scalar field system (dashed) and the corresponding solution of the resonant approximation (solid) for decreasing amplitudes of initial data $\varepsilon$ and increasing truncation $N$ respectively. The left and right panels show data for initial conditions \eqref{eq:20.12_01} and \eqref{eq:20.12_01b}, respectively.}
	\label{fig:RicciOriginEinsteinVSResonant}
\end{figure}

We remark that if $\beta>3/2$ were the exponent of the asymptotic solution, then the prediction that the phases remain finite at $\tau_{*}$, as follows from Eqs. \eqref{eq:22.09.05_02} and \eqref{eq:22.09.05_03} (and also Eq.~\eqref{eq:22.03.21_08} as the special case for $\beta=8/5$), would imply that the resonant approximation should be valid up to the time of the AH formation. Thus, there is no contradiction as in AdS$_{5}$, where the unbounded frequency growth is in tension with the resonant approximation,\footnote{In addition to the approach presented here, the resonant system can also be derived by time averaging \cite{Bizoń.2015}.} despite the fact that the numerical data obtained from the approximation agreed with the nonlinear results \cite{Bizoń.2015}. The hypothesis that the perturbative approach provides a good approximation up to the time of black hole horizon formation is supported in Fig.~\ref{fig:RicciOriginEinsteinVSResonant}, where we compare the resonant approximation with the solution of the Einstein-Scalar field system. Similarly to the AdS$_{4}$ case \cite{Deppe.20191vp}, the convergence with increasing truncation $N$ is relatively slow compared to AdS in higher dimensions. However, the curves appear to approach a limiting solution that agrees with the rescaled solution of \eqref{eq:20.12.18_02}-\eqref{eq:20.12.18_04} in the limit $\varepsilon\rightarrow 0$. This suggests that generic initial data leads to gravitational collapse on the timescale $\varepsilon^{-2}$ as initially reported in \cite{Maliborski.2012}.

\section{Conclusions}
\label{sec:Conclusions}

We provide a strong argument that the \textit{box Minkowski model} \cite{Maliborski.2012} with Dirichlet boundary condition is unstable toward black hole formation for arbitrarily small generic perturbations. We demonstrated this by studying the resonant system for the model and finding that it has a solution that becomes singular in finite time. Our work strengthens the argument that the role of the cosmological constant in the instability of AdS is purely kinematical. Other models with confinement and a resonant spectrum of linear perturbations may also develop a turbulent instability.

Even in the presence of extra resonant terms and in the absence of symmetries in the interaction coefficients, the singular evolution of the system studied here shares features with the respective solution observed in AdS \cite{Bizoń.2015,Deppe.20191vp}. In fact, we demonstrated that the singular solution is determined mainly by the $++-$ resonances.

Although we provide clear evidence for solutions of the resonant system starting with generic initial data that develop a singularity in finite time, the precise nature of the singularity remains unresolved. This is due to the particular feature of $3+1$ dimensional gravitating systems, e.g. in the AdS$_{4}$ case \cite{Deppe.20191vp}, which causes slow convergence of the solutions of the truncated resonant system to the respective solution of the infinite set of equations. However, our results indicate that the singular solution is characterized by the polynomial spectrum of mode amplitudes \eqref{eq:22.07.23_01} with the exponent $\beta$ being close to $8/5$, the value of which agrees with the exponent of the spectrum of mode energies found in the solution of the Einstein-Scalar field system \eqref{eq:22.07.12_04}. Moreover, contrary to the AdS in five and higher spacetime dimensions, where the solution develops an oscillating singularity, we find that here, phases likely stay finite at the singularity. However their higher order derivatives blow up independently of the residual gauge freedom, cf. \cite{Deppe.20191vp}.

An updated and more efficient numerical code allowed us to study much larger resonant systems and to get more precise numerical data. As a result, we significantly improved on previous works \cite{Bizoń.2015, Deppe.20191vp}. However, our efforts to solve large resonant systems have reached current hardware limits. Therefore, we anticipate that follow-up studies will require developing more efficient techniques to reduce the complexity of numerical algorithms which are used to solve resonant systems.

Given our results, the \textit{box Minkowski model} \cite{Maliborski.2012} could prove attractive for further analysis. To the best of our knowledge, this gravitating model has the simplest closed form of interaction coefficients among the models which manifest turbulent instability. Thus it offers an attractive playground for future rigorous development. We hope this work is an essential step toward a better understanding of the instability of AdS$_{4}$ and other resonant systems \cite{Evnin.2021v}.

\vspace{4ex}
\noindent\textit{Acknowledgement.} We thank Anxo Biasi, Piotr Bizo\'n, and Andrzej Rostworowski for valuable comments on earlier version of the manuscript. M.M. acknowledges the support of the Austrian Science Fund (FWF), Project P 29517-N27 and the START-Project Y963. The computational results presented have been achieved [in part] using the Vienna Scientific Cluster (VSC).

\appendix
\section{Boundary time gauge}
\label{sec:BoundaryTimeGauge}

\subsubsection{Resonant system}

In the derivation of \eqref{eq:21.01.08_06}, we have implicitly used the gauge condition $\delta(t,0)=0$. However, \eqref{eq:20.12.18_02} allows us to redefine $\delta(t,r) \rightarrow \delta(t,r) + f(t)$. In particular, we can choose $f(t)$ in a way that gives $\delta(t,1)=0$, so that $t$ is the proper time of the observer located at $r=1$. We redo our calculation of the resonant system in this gauge and get the equation 
\begin{equation}
	\label{eq:21.01.11_01}
	\delta_{2}' = -r \left( \phi_{1}'(t,r)^2 + \dot{\phi}_{1}(t,r)^2 \right)
	\,,
\end{equation}
at second order of $\varepsilon$. Its solution can be written as
\begin{equation}
	\label{eq:21.01.11_02}
	\delta_{2}(t,r) = - \int_{0}^{r} \diff{s} s \left(\phi_{1}'(t,s)^{2} + \dot{\phi}_{1}(t,s)^{2}\right) + f(t)\,.
\end{equation}
We require that $\delta_{2}(t,1)=0$, in agreement with the gauge condition $\delta(t,1)=0$. This yields the condition
\begin{equation}
	\label{eq:21.01.11_03}
	f(t) = \int_{0}^{1} \diff{s} s \left(\phi_{1}'(t,s)^{2} + \dot{\phi}_{1}(t,s)^{2}\right)\,.
\end{equation}
With this, the solution of \eqref{eq:21.01.11_01} can be written as
\begin{equation}
	\label{eq:21.01.11_04}
	\delta_{2}(t,r) = \int_{r}^{1} \diff{s} s \left(\phi_{1}'(t,s)^{2} + \dot{\phi}_{1}(t,s)^{2}\right)\,.
\end{equation}
For the analogous calculation for AdS, see \cite{Craps.2015}. Repeating our calculations \eqref{eq:20.12.18_32}-\eqref{eq:20.12.18_36}, we find
\begin{align}
	\label{eq:21.01.11_05}
	\inner{\dot{\delta}_{2}\dot{\phi}_{1}}{e_{l}} &= \int_{0}^{1} \diff{r} r^{2} e_{l}(r) \dot{\phi}_{1}(t,r) \int_{r}^{1} \diff{s} s \left(\frac{\partial}{\partial t} \phi_{1}'(t,s)^{2} + \frac{\partial}{\partial t} \dot{\phi}_{1}(t,s)^{2}\right) \notag \\
	&= \sum_{i,j,k=0}^{\infty} \dot{c}_{k}(t) \int_{0}^{1} \diff{r} r^{2} e_{l}(r) e_{k}(r) \notag \\
	&\quad \int_{r}^{1} \diff{s} s \left[\frac{\partial}{\partial t} \left(c_{i}(t) c_{j}(t) \right)e_{i}'(s) e_{j}'(s) +  \frac{\partial}{\partial t} \left(\dot{c}_{i}(t) \dot{c}_{j}(t) \right)e_{i}(s) e_{j}(s)\right]\,,
\end{align}
\begin{align}
	\label{eq:21.01.11_06}
	\inner{\delta_{2}\ddot{\phi}_{1}}{e_{l}} &= \int_{0}^{1} \diff{r} r^{2} e_{l}(r) \ddot{\phi}_{1}(t,r) \int_{r}^{1} \diff{s} s \left(\phi_{1}'(t,s)^{2} + \dot{\phi}_{1}(t,s)^{2}\right) \notag \\
	&= \sum_{i,j,k=0}^{\infty} \ddot{c}_{k}(t) \int_{0}^{1} \diff{r} r^{2} e_{l}(r) e_{k}(r) \notag \\
	&\quad \int_{r}^{1} \diff{s} s \left[ c_{i}(t) c_{j}(t) e_{i}'(s) e_{j}'(s)  + \dot{c}_{i}(t) \dot{c}_{j}(t) e_{i}(s) e_{j}(s)\right]\,.
\end{align}
The rest of the calculations \eqref{eq:20.12.18_32}-\eqref{eq:20.12.18_36} stays the same. Defining the integrals
\begin{align}
	\label{eq:21.01.11_07}
	N^{*}_{klij} &= - \int_{0}^{1}\diff{r} r^{2} e_{l}e_{k}\int_{r}^{1}\diff{s}s e_{i}'e_{j}'\,,
	\\
	\label{eq:21.01.11_08}
	N_{klij} &= - \int_{0}^{1}\diff{r} r^{2} e_{l}e_{k}\int_{r}^{1}\diff{s}s e_{i}e_{j}\,,
\end{align}
we get the resonant system
\begin{equation}
	\label{eq:21.01.11_09}
	2i\omega_{l}\frac{\diff{} \alpha_{l}}{\diff{\tau}} = \sum_{ijk}^{+++} \tilde{R}_{klij} \alpha_{i} \alpha_{j} \alpha_{k} + \sum_{ijk}^{-++} \tilde{S}_{klij} \bar{\alpha}_{i} \alpha_{j} \alpha_{k} + \sum_{ijk}^{+--} \tilde{S}_{klij} \alpha_{i} \bar{\alpha}_{j} \bar{\alpha}_{k}\,,
\end{equation}
where $\tilde{S}_{klij} = \tilde{O}_{klij} + \tilde{Q}_{klji} + \tilde{P}_{ilkj}$ and $\tilde{O}_{klij}$, $\tilde{Q}_{klij}$, and $\tilde{P}_{klij}$ are as in \eqref{eq:21.08.02_04}-\eqref{eq:21.08.02_06}, with
\begin{equation}
	L^{(*)}_{klij} \rightarrow N^{(*)}_{klij}\,.
\end{equation}

We can express $N_{klij}$ in terms of $L_{klij}$ by combining the respective definitions to get
\begin{equation}
	\label{eq:22.03.16_32}
	L_{klij} - N_{klij} = \int_{0}^{1}\diff{r}r^{2}e_{k}e_{l}\int_{0}^{1}\diff{s}s e_{i}e_{j}\,,
\end{equation}
similarly for $N^{*}_{klij}$. By explicit integration, we find
\begin{equation}
	\label{eq:22.03.16_33}
	N_{klij} = L_{klij} - \delta_{kl}\left(-\Ci(\omega_{i}+\omega_{j}) + \log(\omega_{i}+\omega_{j}) + \Ci(|\omega_{i}-\omega_{j}|) -\log|\omega_{i}-\omega_{j}| \right)\,,
\end{equation}
and
\begin{multline}
	\label{eq:22.03.16_34}
	N^{*}_{klij} = L^{*}_{klij}+
	\\
	- \delta_{kl}\left[ \frac{1}{2}\left(\omega_{i}^{2}+\omega_{j}^{2}\right)\left(-\Ci(\omega_{i}+\omega_{j}) + \log(\omega_{i}+\omega_{j}) + \Ci(|\omega_{i}-\omega_{j}|) -\log|\omega_{i}-\omega_{j}|\right) - \omega_{i}\omega_{j} \right]\,.
\end{multline}
The $i=j$ cases in \eqref{eq:22.03.16_33}-\eqref{eq:22.03.16_34} are obtained by taking the limit of the corresponding expressions.

\subsubsection{Interaction coefficients}
The corresponding expressions for \eqref{eq:22.03.15_03} in the boundary time gauge are
\begin{equation}
	\label{eq:22.06.24_01}
	T_{i} = 2 \omega _i^3 \left(-10 \text{Si}\left(2 \omega _i\right)+5 \text{Si}\left(4
   \omega _i\right)+8 \omega _i\right)
   \,,
\end{equation}
and
\begin{multline}
	\label{eq:22.06.24_02}
	R_{il} =  16 \omega _i^2 \omega _l^2
	\\
	+\frac{8 \omega _l \left(2 \omega _l^4-\left(\omega _i^2-\omega
   _l^2\right){}^2\right) \text{Si}\left(2 \omega _l\right)}{\omega
   _i^2-\omega _l^2}-\frac{8 \omega _i \left(2 \omega _i^4-\left(\omega
   _i^2-\omega _l^2\right){}^2\right) \text{Si}\left(2 \omega
   _i\right)}{\omega _i^2-\omega _l^2}
   \\
   +\frac{2 \left(\left(\omega _i-\omega
   _l\right){}^4+\left(\omega _i^2+\omega _l^2\right){}^2\right)
   \text{Si}\left(2 \left(\omega _i-\omega _l\right)\right)}{\omega _i-\omega
   _l}
   \\
   +\frac{2 \left(\left(\omega _i+\omega _l\right){}^4+\left(\omega
   _i^2+\omega _l^2\right){}^2\right) \text{Si}\left(2 \left(\omega _i+\omega
   _l\right)\right)}{\omega _i+\omega _l}
   \,,
\end{multline}
cf. \eqref{eq:22.06.06_01}-\eqref{eq:22.06.06_02}, while for $k=i+j-l$ with $i\neq l \wedge j\neq l$ the coefficients $S^{++-}_{ijkl}$ remain unchanged, see \eqref{eq:22.06.06_03}. For the $+++$ and $+--$ resonant terms, the gauge contribution cancels out, hence the expression \eqref{eq:22.06.06_04} and \eqref{eq:22.06.06_05} remain valid for the boundary time gauge.

\subsubsection{Conserved quantities}
Note that in this gauge, we have
\begin{equation}
	R_{ij}=R_{ji}
	\,,
\end{equation}
so $R^{A}_{ij}=0$, see Eq.~\eqref{eq:22.03.16_06}, and $\tilde{V}_{j}\equiv 0$, cf. \eqref{eq:22.03.16_10}, thus the corresponding conserved quantity $H$ is simply
\begin{equation}
	H = W = X + \frac{1}{2}V + \frac{1}{3}Z
	\,,
\end{equation}
for the full system
and 
\begin{equation}
	H = \frac{1}{2}V
	\,,
\end{equation}
for the $++-$ system, cf. the respective expressions \eqref{eq:22.03.16_13} and \eqref{eq:22.03.16_31}. The other known integrals of motion, $E$ and $J$ (for the $++-$ system only), are not affected by the residual gauge choice.

Since, in this case, the flow is Hamiltonian, cf. \eqref{eq:22.03.16_08} and \eqref{eq:22.03.16_28}, these conserved quantities are the Noether charges which follow from the respective symmetries \eqref{eq:22.08.31_01}, as mentioned in the main text.

\subsubsection{Numerical solution}

\begin{figure}[!t]
	\includegraphics[width=0.495\textwidth]{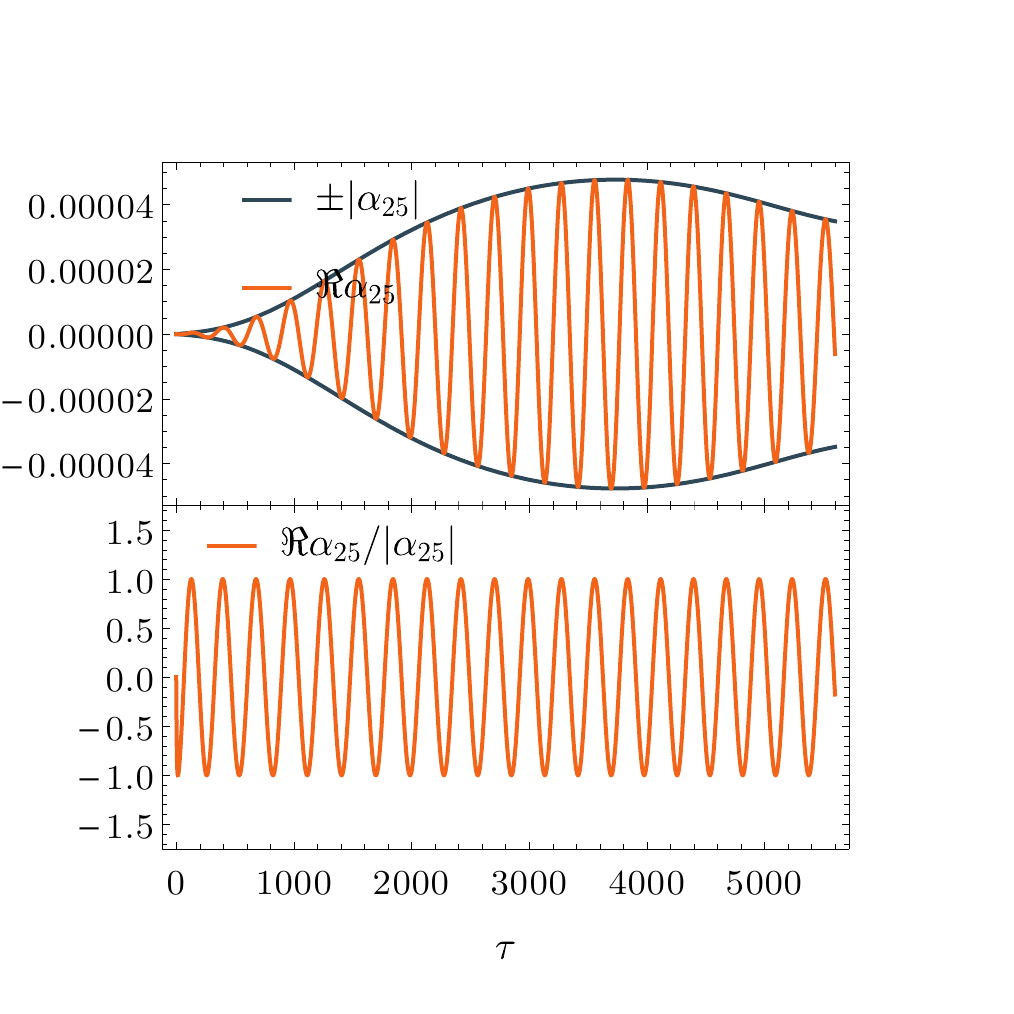}
	\includegraphics[width=0.495\textwidth]{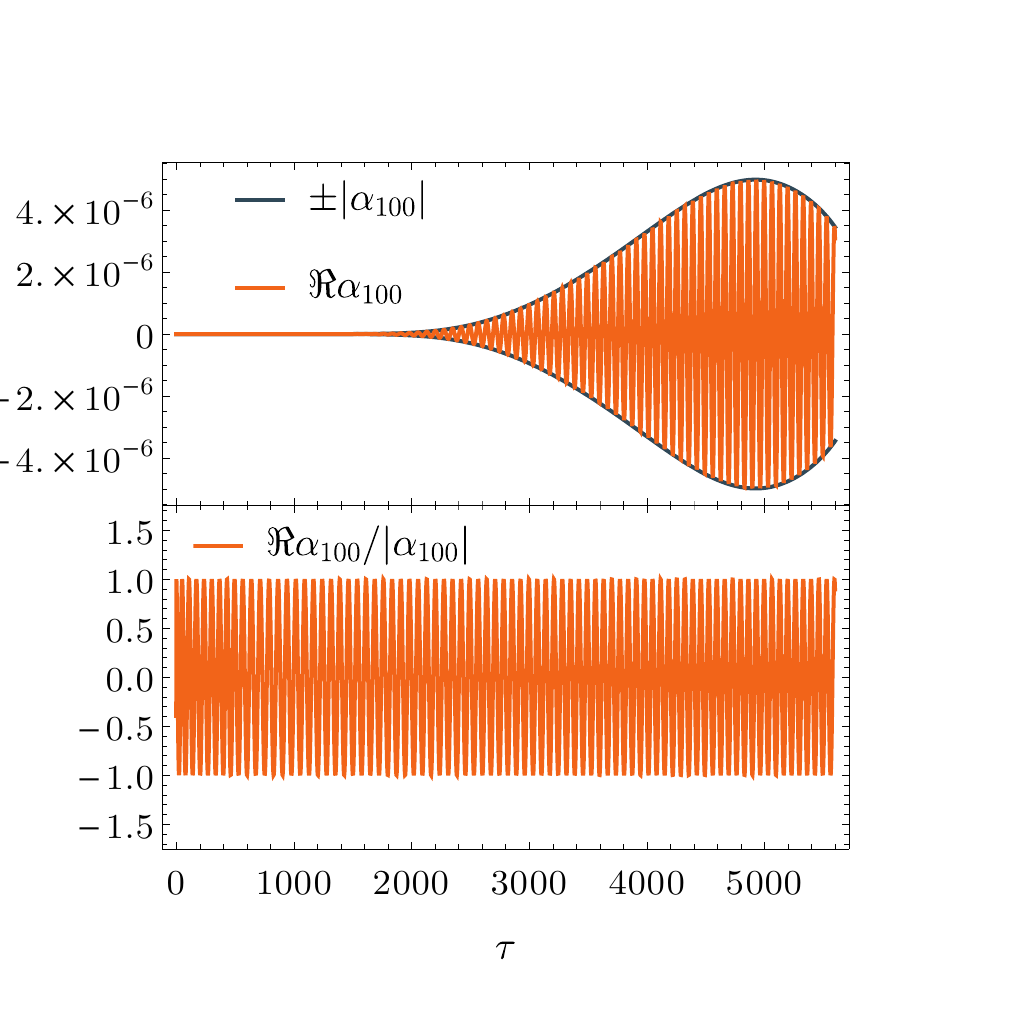}
	\caption{Time evolution of sample modes in the boundary time gauge. Note the higher frequency of the corresponding modes compared to the origin gauge case presented in Fig.~\ref{fig:RS_AmplitudeAndRe_Set1}.}
	\label{fig:RS_AmplitudeAndRe_Set1_BND}
\end{figure}

\begin{figure}[!t]
	\includegraphics[width=0.4\textwidth]{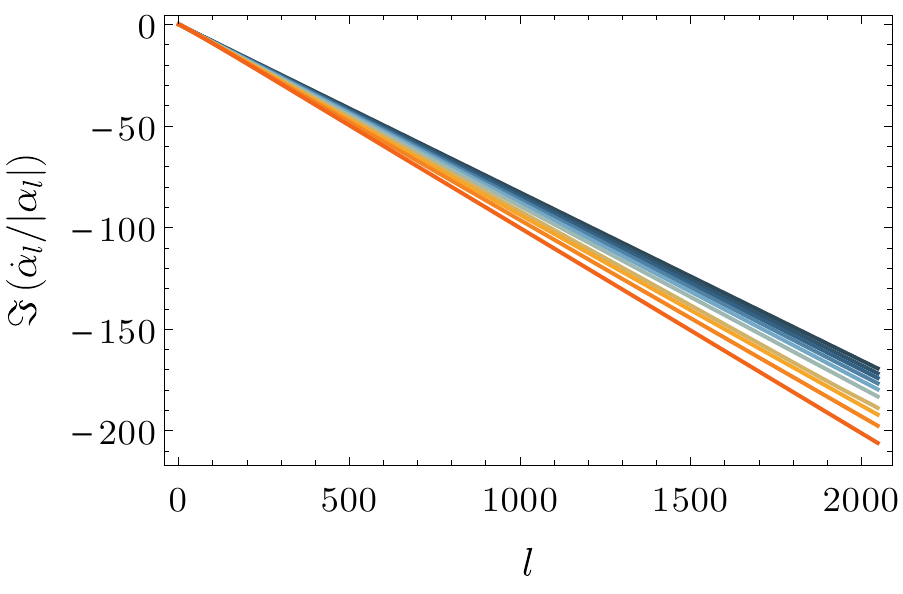}
	\caption{Evolution of the phase derivatives in the boundary time gauge. As in Fig.~\ref{fig:RSAmplitudePhaseEvolution} time is color coded and increases from bluish to reddish colors.}
	\label{fig:RSAmplitudePhaseEvolution_BND}
\end{figure}

Interestingly, as for the AdS case, we see that the evolution of the amplitudes is independent of the gauge choice, so that the amplitudes found in the origin gauge agree (up to truncation errors) with the amplitudes determined in the boundary gauge. Therefore, we also observe a singular solution here. However, the phases differ, compare Fig.~\ref{fig:RS_AmplitudeAndRe_Set1} and Fig.~\ref{fig:RS_AmplitudeAndRe_Set1_BND}, which is the consequence of the distinct coefficients \eqref{eq:22.06.24_01}-\eqref{eq:22.06.24_02} in Eq.~\eqref{eq:22.06.28_03}. As a result, each mode's oscillation frequency is higher in the boundary time gauge than in the origin time gauge. However, if the oscillation frequency increases as the solution approaches the singularity, the growth is less noticeable in this time gauge. Regardless, the phases stay synchronised during the evolution; see Fig.~\ref{fig:RSAmplitudePhaseEvolution_BND}.

\begin{figure}
	\includegraphics[width=0.9\textwidth]{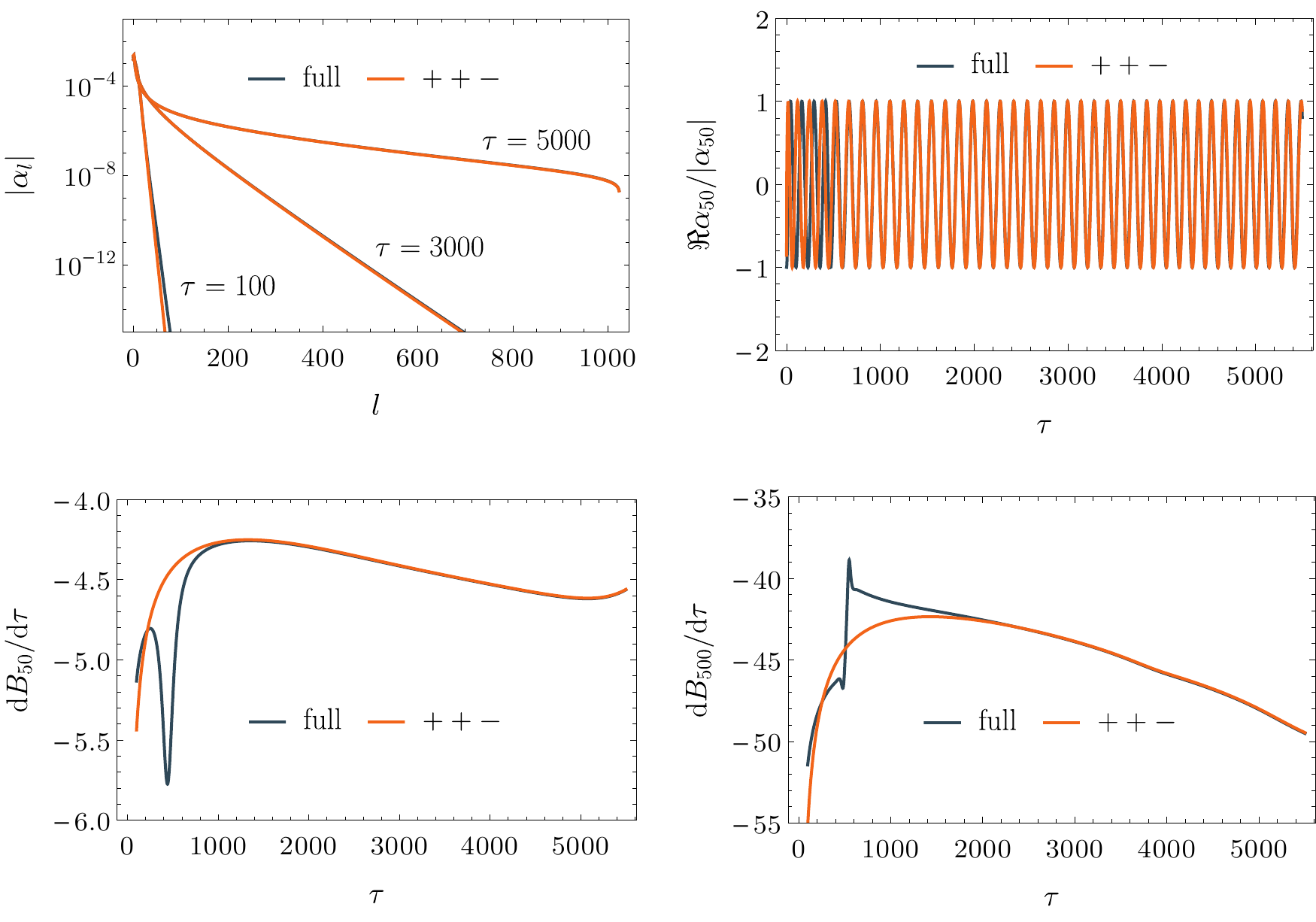}
	\caption{Analog of Fig.~\ref{fig:RS_PPPvsPPMvsPMM2_Set1} for the boundary time gauge.}
	\label{fig:RS_PPPvsPPMvsPMM2_Set1_BND}
\end{figure}

As above, we find that the $++-$ resonant term dominates the evolution. The comparison of solutions of the full and the $++-$ resonant systems starting from the same initial conditions is presented in Fig.~\ref{fig:RS_PPPvsPPMvsPMM2_Set1_BND}. The apparent agreement between the solutions strongly suggests that the $++-$ resonances largely determine the singular solution.

\subsubsection{Asymptotic solution}
We follow the same steps as in Sec.~\ref{sec:AsymptoticSolution} for the analysis in the origin gauge. The time evolution of the amplitudes is almost identical to the behavior in the previous case, cf. Fig.~\ref{fig:RS_PPPvsPPMvsPMM2_Set1} and Fig.~\ref{fig:RS_PPPvsPPMvsPMM2_Set1_BND}, so the fits of the ansatz \eqref{eq:22.07.23_01} give similar values for the exponent $\beta$ and the time $\tau_{*}$ when the analyticity strip radius crosses zero. (Fits to the data with $N=2048$ mode truncation give $\beta\approx 1.53$ and $\tau_{*}\approx 5467$. In this case, we see a smaller variation with respect to the fitting interval.)

\begin{figure}[!t]
	\includegraphics[width=0.99\textwidth]{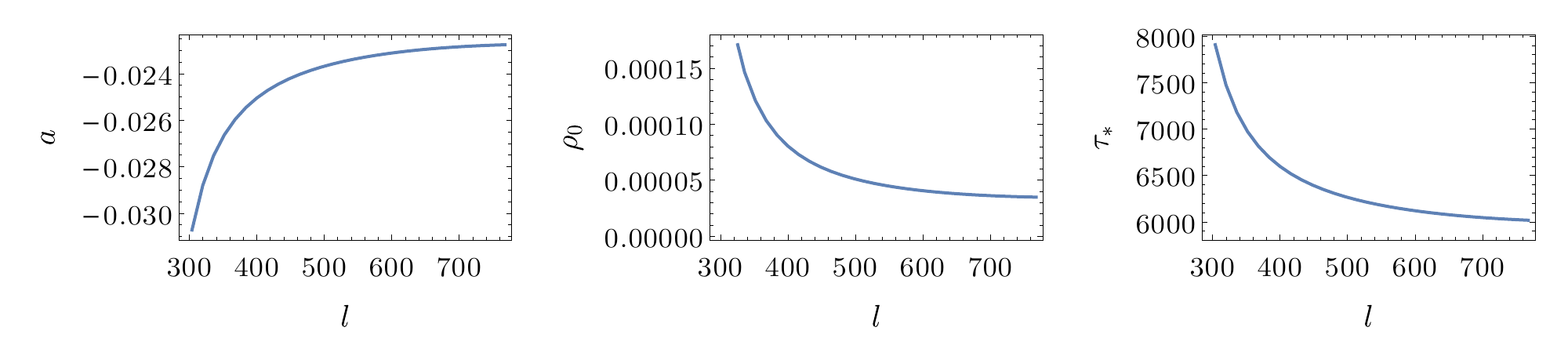}
	\caption{The dependence of the fitting parameters in \eqref{eq:22.07.21_02} on the mode number $l$. In the asymptotic regime when $\tau\approx\tau_{*}$ and $l\rightarrow\infty$, the fitting parameters should be independent of $l$.}
	\label{fig:fitAvsB_2_BND}
\end{figure}

The leading order expansion of $R_{il}$ lacks the logarithmic terms present in \eqref{eq:22.09.05_01}, thus the analog of \eqref{eq:22.09.05_02} in the boundary time gauge is
\begin{equation}
	\label{eq:22.09.05_06}
	\frac{\diff{B_{l}}}{\diff{\tau}} \approx l \left[
	c_{2} \text{Li}_{2\beta-2}\left(e^{-2\rho}\right)
	+ c_{4} \text{Li}_{2\beta-1}\left(e^{-2\rho}\right)
	\right]
	\,.
\end{equation}
Thus although higher order derivatives blow up for $\beta\leq 3/2$ and $\beta>3/2$, the first derivative stays finite at $\tau_*$ for $\beta>3/2$ only. Fitting \eqref{eq:22.09.05_06} to the numerical data was not successful. Therefore, as before, we fixed $\beta=8/5$ and fitted the corresponding asymptotic formula
\begin{equation}
	\label{eq:22.07.21_02}
	\frac{\diff{B_{l}}}{\diff{\tau}} \approx a\,l\left( 1 + b_{1}\rho^{1/5} \right)
	\,,
	\quad \Rightarrow \quad 
	\frac{\diff{}^{2}B_{l}}{\diff{\tau}^{2}} \sim \rho^{-4/5}
	\,,
\end{equation}
with $a$ and $b_{1}$ constant. We find that \eqref{eq:22.07.21_02} matches the data well. Repeating the fit for various mode numbers $l$, we observe the convergence of the fitting parameters for $l\rightarrow\infty$, see Fig.~\ref{fig:fitAvsB_2_BND}.
Surprisingly, although the value of $\tau_{*}$ from the analysis of the spectrum of amplitudes agrees with the value computed by solving the Einstein-Scalar field system, the estimate obtained from the analysis of the phases differs.

Also in this case, the resonant approximation accurately reproduces the Ricci scalar evaluated at $r=0$, and the limiting solution is approached when the truncation $N$ is increased. We omit the plot demonstrating this as it appears remarkably similar to the one presented in Fig.~\ref{fig:RicciOriginEinsteinVSResonant}.

\printbibliography

\end{document}